\documentclass[twocolumn, secnumarabic,amssymb, nobibnotes, aps, prb,groupedaddress,superscriptaddress]{revtex4-2}
\usepackage{amsmath,graphicx,latexsym,times,color}
\usepackage{setspace}
\usepackage[hidelinks,colorlinks=true, allcolors=blue]{hyperref}
\usepackage{array}
\usepackage{textcomp}
\usepackage{titlesec}
\usepackage{physics}
\usepackage{gensymb}
\usepackage{nccmath}
\usepackage{empheq} %% loads mathtools, which loads amsmath
\usepackage[normalem]{ulem} % for \sout{} command

\definecolor{blue-violet}{rgb}{0.54, 0.17, 0.89} %Vector

\let\oldtimes\times  % Make the times "x" use less spacing
\renewcommand\times{{\oldtimes}}

\usepackage[cmyk,dvipsnames]{xcolor}
\definecolor{darkorchid}{HTML}{bf3eff}
\usepackage[cmyk,dvipsnames]{xcolor}
\definecolor{darkorchid}{HTML}{bf3eff}

\begin{document}

\title{Beating micromagnetic limits on skyrmion stability by long-range frustration}

\author{Shiwei Zhu}
\affiliation{Zhejiang Key Laboratory of Quantum State Control and Optical Field Manipulation and Department of Physics,
\href{https://ror.org/03893we55}{Zhejiang Sci-Tech University}, Hangzhou 310018, China}
\affiliation{\href{https://ror.org/01ahyrz84}{Universit\'e de Toulouse}, \href{https://ror.org/02feahw73}{CNRS}, \href{https://ror.org/03kwnqq69}{CEMES}, Toulouse, France}

\author{Moritz A. Goerzen}
\affiliation{\href{https://ror.org/01ahyrz84}{Universit\'e de Toulouse}, \href{https://ror.org/02feahw73}{CNRS}, \href{https://ror.org/03kwnqq69}{CEMES}, Toulouse, France}

\author{Changsheng Song}
\email[Contact author: ]{cssong@zstu.edu.cn}
\affiliation{Zhejiang Key Laboratory of Quantum State Control and Optical Field Manipulation and Department of Physics,
\href{https://ror.org/03893we55}{Zhejiang Sci-Tech University}, Hangzhou 310018, China}

\author{Dongzhe Li}
\email[Contact author: ]{dongzhe.li@cemes.fr}
\affiliation{\href{https://ror.org/01ahyrz84}{Universit\'e de Toulouse}, \href{https://ror.org/02feahw73}{CNRS}, \href{https://ror.org/03kwnqq69}{CEMES}, Toulouse, France}	
	
\date{\today}
	
\begin{abstract}

Skyrmion stability is commonly assumed to scale with skyrmion size or exchange stiffness within micromagnetic models. Here, we demonstrate that long-range exchange frustration can break this paradigm, enhancing the collapse energy barrier without increasing skyrmion size or magnetic energy scale. By mapping the continuum model onto a spin-lattice Hamiltonian, we find that skyrmions with identical micromagnetic parameters can exhibit significantly different energy barriers, depending on their underlying atomistic exchange interactions. We attribute this behavior to saddle point textures, whose pronounced noncollinearity captures long-range frustration beyond the micromagnetic approximation. We further develop an exchange optimization framework to predict that long-range frustration can double the energy barrier in physically realistic conditions, possibly valid for ultrathin films or van der Waals magnets. 
These results hold across different lattice symmetries, revealing an intrinsic limitation of micromagnetics and establishing long-range frustration engineering as a promising route toward highly stable nanoscale skyrmions.
%Our predictions hold for different lattice symmetries. Our results reveal an intrinsic limitation of micromagnetics and establish long-range frustration engineering as a promising route toward highly stable nanoscale skyrmions.

\end{abstract}
	
\maketitle

\begin{figure*}[tbp]%---------------------------------------------------------------------------------------------------------------------Figure 1
	\centering
	\includegraphics[width=0.95\linewidth]{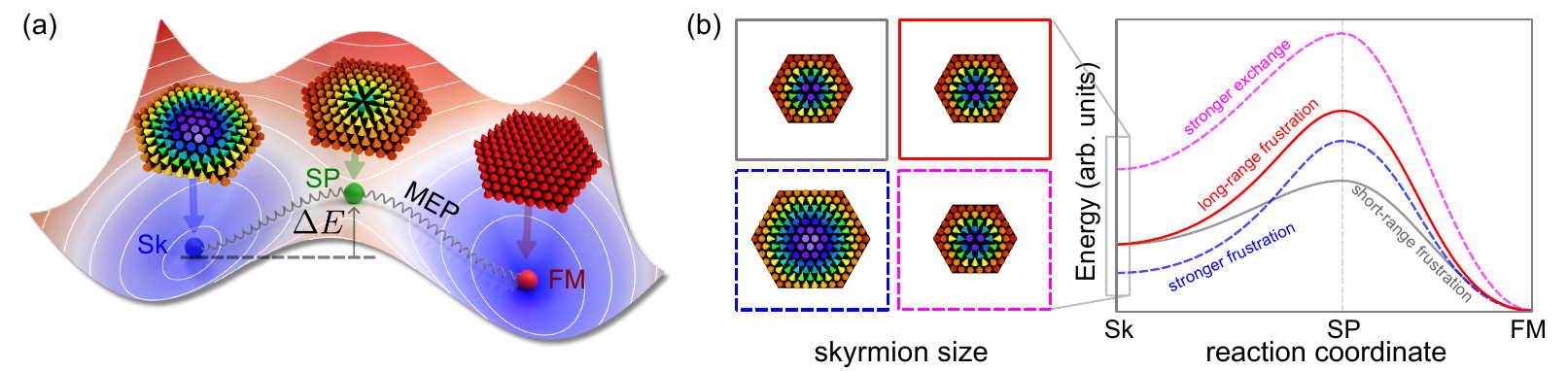}
	\caption{\label{fig_1}
    Schematic of the central mechanism.
	(a) Along the minimum energy path (MEP), the skyrmion (Sk) transitions into the ferromagnetic (FM) state after crossing the saddle point (SP). 
	(b) The MEPs of different approaches to increasing the energy barrier are shown in comparison to the original case with short-range frustration (grey), along with their corresponding skyrmion sizes. Long-range exchange frustration (red) increases the energy barrier $\Delta E$ while preserving skyrmion size and energy scale, in contrast to conventional approaches based on increasing short-range frustration strength (blue dashed) or exchange (violet dashed).
}
\end{figure*}

Topological spin textures play a central role in modern magnetism, attracting great interest for both fundamental studies and spintronic applications. 
Among them, magnetic skyrmions are topologically protected quasiparticles characterized by a whirling spin texture in real space. Their nanoscale size~\cite{heinze2011spontaneous,Romming2013,Dongzhe2022_fgt}, combined with good thermal stability, efficient electrical manipulation and detection \cite{fert2013,sampaio2013nucleation,urrestarazu2024electrical,li2024proposal,chen2024all}, and ultrafast optical control~\cite{buttner2021observation,dabrowski2022all,khela2023laser}, makes them particularly promising for next-generation spintronic devices~\cite{fert2013skyrmions,fert2017magnetic,back20202020,gobel2021beyond}. A key requirement for such applications is skyrmion stability, characterized by the annihilation energy barrier $\Delta E$ between the skyrmion state and the saddle point (SP) along the minimum energy path (MEP) to a ferromagnetic (FM) state [Fig.~\ref{fig_1}(a)]. This barrier originates from competing magnetic interactions, including the Dzyaloshinskii-Moriya interaction (DMI), dipolar interactions, higher-order exchange interactions, and frustrated exchange interactions~\cite{PhysRevB.101.045416, paul2020role, von2017enhanced}, which stabilize noncollinear spin textures against ferromagnetic alignment. Although there are also alternative routes such as entropic stabilization~\cite{varentcova2020toward}, realizing room-temperature skyrmions still requires effective strategies to directly enhance $\Delta E$.

%-----------------------------------------------------------------------------------------------------------------------------

A conventional approach is to strengthen competing interactions within a suitable range. However, this typically increases skyrmion size [e.g., dashed blue line in Fig.~\ref{fig_1}(b)], leading to the widely observed trend that larger skyrmions exhibit higher stability~\cite{varentsova2018interplay,Soum_2018,megha2025,Soum2026}. 
%\shiwei{Achieving simultaneously small size and large $\Delta E$ therefore remains a central challenge for applications.}
Breaking the dilemma between small size and large $\Delta E$ is a necessary prerequisite since practical applications require nanoscale skyrmions that remain robust against thermal fluctuations.
An alternative strategy is to employ materials with stronger exchange, which increases the overall micromagnetic energy scale and thereby enhances $\Delta E$ without significantly changing skyrmion size [Fig.~\ref{fig_1}(b) violet dashed], consistent with the conventional micromagnetic picture where the barrier scales with exchange stiffness~\cite{Masell2019}. 
However, it globally elevates the magnetic energy landscape, making skyrmions more difficult to generate and manipulate under realistic external stimuli.
These limitations motivate the search
for a mechanism that can selectively elevate $\Delta E$ while decoupling it from both skyrmion size and magnetic energy scale.

% Among these competing interactions, exchange frustration is particularly distinctive because it can emerge from oscillations inherent to the Ruderman–Kittel–Kasuya–Yosida (RKKY) mechanism~\cite{Yosida1996,Lit1998,Bruno1995,Hayami2016}. 
% On a discrete lattice of spins $\mathbf{m}_i$, these oscillations
% %can be captured in an isotropic Heisenberg model
% \shiwei{are encoded in the shell-resolved exchange vector $\mathbf{J}=(J_1,\ldots,J_S)$ of the isotropic Heisenberg model}
% \begin{equation}\label{eq:exchange_hamiltonian}
% \mathcal{H} = -\sum_{s=1}^{S} J_{s} \sum_{\langle i,j\rangle_s} \mathbf{m}_i \cdot \mathbf{m}_j~,
% \end{equation}
% % where they manifest as long-range (LR) modulations of shell resolved constants $\mathbf{J}=(J_1,...,J_S)$, 
% \shiwei{giving rise to long-range (LR) frustrated interactions 
% % which are 
% known to stabilize a wide variety of topological spin textures~\cite{Okubo2012,leonov2015multiply,dupe2016engineering,von2017enhanced,rybakov2022magnetic,sallermann2023stability}.} 
%Among these competing interactions, exchange frustration is particularly distinctive because it can emerge from oscillations inherent to the Ruderman–Kittel–Kasuya–Yosida (RKKY) mechanism~\cite{Yosida1996,Lit1998,Bruno1995,Hayami2016}. 
%On a discrete lattice of spins $\mathbf{m}_i$, these oscillations

Exchange frustration is known to stabilize a wide variety of topological spin textures~\cite{Okubo2012,leonov2015multiply,dupe2016engineering,von2017enhanced,rybakov2022magnetic,sallermann2023stability}. On a discrete lattice of spins $\mathbf{m}_i$, exchange interactions are described by the isotropic Heisenberg model
\begin{equation}\label{eq:exchange_hamiltonian}
\mathcal{H} = -\sum_{s=1}^{S} J_s \sum_{\langle i,j\rangle_s} \mathbf{m}_i \cdot \mathbf{m}_j.
\end{equation}
Unlike other interactions, exchange frustration can emerge from the Ruderman–Kittel–Kasuya–Yosida (RKKY) mechanism~\cite{Yosida1996,Lit1998,Bruno1995,Hayami2016}, giving rise to long-range (LR) exchange interactions.
However, previous studies were predominantly confined to simplified short-range (SR) models~\cite{zhang2017skyrmion,varentsova2018interplay,Masell2019, desplat2019paths,Zhang2021} limited to a few exchange shells (e.g., $J_1 - J_2$ or $J_1 - J_2 - J_3$).

In this Letter, we investigate the fundamental boundaries that limit the stability of sub-10 nm skyrmions in frustrated 2D magnets. Based on an advanced micromagnetic model \cite{rybakov2022magnetic}, we 
develop a procedure that enables us to parameterize Eq.~(\ref{eq:exchange_hamiltonian}) in various ways, by keeping the skyrmion radius nearly constant. From subsequent geodesic nudged elastic band (GNEB) \cite{bessarab2015method} calculations, we find $\Delta E$ varies for different atomistic realizations.
Exploiting this discrepancy, we discover that LR exchange frustration is a superior mechanism for breaking micromagnetic limits, as shown in Fig.~\ref{fig_1}(b). We further develop an optimization framework that systematically explores the $S$-dimensional exchange space and yields a general description of the optimal barrier. This advantage of LR frustration is universal, as demonstrated on square, hexagonal, and honeycomb lattices.

\textit{Continuum-discrete mapping.}
We establish the correspondence between the continuum and discrete lattice models following Refs~\cite{rybakov2022magnetic,sallermann2023stability}. In two dimensions, the energy can be approximated in terms of a fourth-order expansion of the magnetization $\mathbf{m(\tilde{\mathbf{r}})}$ 
\begin{equation}\label{eq:micromagnetic_model}
\begin{aligned}
\mathcal E=\int_{\mathbb{R}^2}\Bigg[
& \sum_{\alpha}
\left(\frac{\partial \mathbf m}{\partial \tilde r_\alpha}\right)^2
 +(1-\gamma)\sum_{\alpha, \beta\neq \alpha}
\left(\frac{\partial^2 \mathbf m}{\partial \tilde r_\alpha^2}
-\frac{\partial^2 \mathbf m}{\partial \tilde r_\beta^2}\right)^2 \\
&+\gamma\sum_{\alpha, \beta\neq \alpha}
\left(\frac{\partial^2 \mathbf m}{\partial \tilde r_\alpha\partial \tilde r_\beta}\right)^2
\Bigg] ~\mathrm{d}^2 \tilde{\mathbf{r}}~,
\end{aligned}
\end{equation}
where $\alpha, \beta \in \{x,y\}$ and $\tilde{\mathbf{r}} = \mathbf{r} / r_0\in\mathbb{R}^2$ is a dimensionless position vector.
The dimensionless energy $\mathcal{E}=E/(Ar_0)$ is parameterized by the reduced constants
\begin{equation}
    r_0 = \sqrt{\frac{B+C}{A}}~,\quad \gamma=\frac{C}{B+C}~,
\end{equation}
which are determined by the discrete exchange vector $\mathbf{J}$ via $A=\mathbf{a}\cdot\mathbf{J}$, $B=\mathbf{b}\cdot\mathbf{J}$ and $C=\mathbf{c}\cdot\mathbf{J}$. 
The coefficients $\mathbf{a},\mathbf{b},\mathbf{c}\in\mathbb{R}^S$ depend directly on the lattice geometry (see Supplemental Material (SM), Sec.~S1 and Table~S1 \cite{supplmat}).
This reduces the high-dimensional space of discrete exchange constants to the micromagnetic parameters $(\gamma, r_0, A)$, 
where $A$ denotes exchange stiffness, $r_0$ sets the intrinsic length scale, and $\gamma$ controls the anisotropy of the fourth-order terms.

\begin{figure*}[t]%---------------------------------------------------------------------------------------------------------------------Figure 2
	\centering
	\includegraphics[width=0.9\linewidth]{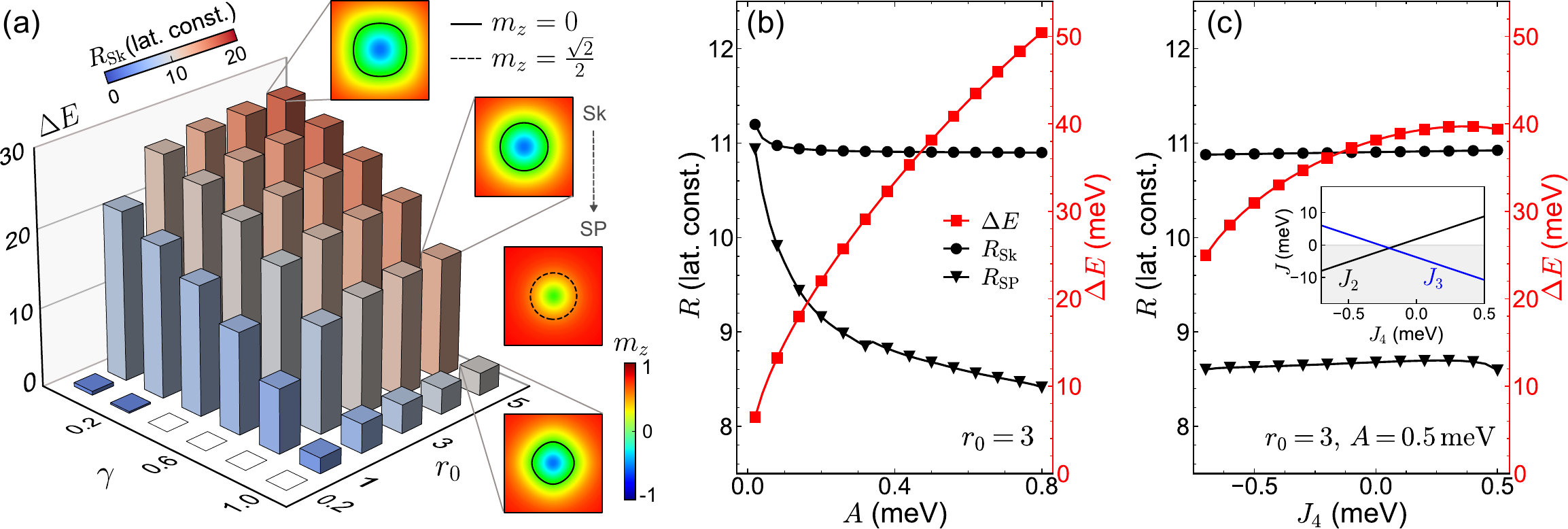}
	\caption{\label{fig_2}
Limitations of micromagnetic models.
(a) Energy barrier $\Delta E$ (bar height) and skyrmion radius $R_{\mathrm{Sk}}$ (color scale, in units of the lattice constant) as functions of micromagnetic parameters $(\gamma, r_0)$. Insets show skyrmion configurations at $r_0=5$ for $\gamma=0$, $0.8$, and $1$, with the corresponding SP at $\gamma=0.8$. Black circles mark $m_z=0$ (solid) and $m_z=\sqrt{2}/2$ (dashed), defining the skyrmion and SP radii, respectively. 
(b) $\Delta E$ (blue stars), $R_\text{Sk}$ (black dots) and SP radius $R_\text{SP}$ (black triangles) as a function of $A$ at fixed $\gamma=0.8$ and $r_0 = 3$. The SP variation gives rise to the non-linear dependence of $\Delta E$ on $A$.
(c) At fixed $\gamma=0.8$, $r_0=3$ and $A=0.5$, $\Delta E$ can be widely tuned by varying $J_4$, while $R_\text{Sk}$ and $R_\text{SP}$ remain nearly constant, in stark contrast to the constant micromagnetic prediction. Inset: The corresponding evolution of other exchange terms, showing a crossover of the dominant frustration term from $J_2$ to $J_3$.
}
\end{figure*}%-----------------------------------------------------------------------------------------------------------------------------

\textit{Skyrmion size vs. stability.}
In $C_6$ lattices (hexagonal and honeycomb), the continuum mapping enforces $C=4B$, fixing $\gamma=0.8$ (see Sec.~S2 in the SM \cite{supplmat}).  This leaves no freedom to study the role of $\gamma$.
We therefore start from a square lattice and identify $\gamma \in [0,1]$ and $r_0 \in [0,5]$ as the region of primary interest based on RKKY-type exchange interaction (see Sec.~S3 in SM~\cite{supplmat}).
%Note that here and t
Throughout this work, $J_1=10~$meV is fixed to keep the nearest-neighbor coupling constant.
%Considering this, 
We first include exchange interactions up to $J_3$ so that the mapping 
$(\gamma, r_0)\to(J_2,J_3)$ is determined (see Sec.~S4 in SM~\cite{supplmat}). 
Obtaining the respective $\mathbf{J}$ for values of $\gamma$ and $r_0$ in the previously identified region of interest, we minimize isolated skyrmions with respect to the Hamiltonian in Eq.~(\ref{eq:exchange_hamiltonian}) and compute the MEP and corresponding $\Delta E$ using GNEB (see Sec.~S5 in SM~\cite{supplmat}) via the \textsc{spinaker} code. We measure the size of skyrmions by radius $R_\text{Sk}$, defined as the distance from the skyrmion center to the contour line with $m_z=0$. On the other hand, the SP radius $R_\text{SP}$ is determined from the contour with $m_z=\sqrt{2}/2$, which corresponds to half of the spin rotation between the core and the FM background.

The results of this analysis are presented in Fig.~\ref{fig_2}(a), where the height of the bars indicates $\Delta E$ and the color represents $R_\text{Sk}$. 
We observe that $R_\text{Sk}$ and $\Delta E$ monotonically rise with $r_0$, where $\Delta E$ shows a gradual saturation for $r_0 \gtrsim 3$. 
For sufficiently small values of $r_0$, the skyrmion loses metastability and collapses into the FM state, corresponding to the blank regions in Fig.~\ref{fig_2}(a). 
Additionally, we observe that the parameter $\gamma$ controls the shape of the skyrmion. As shown in Fig.~\ref{fig_2}(a), the shape evolves from a square-like form at $\gamma=0$ to a diamond-like form at $\gamma=1$, crossing the radially symmetric skyrmion at $\gamma=0.8$. Because $\gamma \propto C$, these deformations are thus absent in studies that neglect $C$ \cite{banik2026paradoxical, Masell2019}. 
Although $\gamma$ also affects $R_\text{Sk}$ and $\Delta E$, we find $\gamma \approx 0.8$ is an accumulation point, even for the square lattice (Fig.~S1 in SM~\cite{supplmat}). To simplify the analysis, we therefore focus on the hexagonal lattice (which always has $\gamma=0.8$) for subsequent calculations.

The exchange stiffness $A$ sets the overall energy scale in the continuum model. Consequently, a linear scaling of $\Delta E$ with $A$ is expected, as reflected in analytical estimations for Belavin-Polyakov skyrmions $\Delta E = 4\pi A - E_{\text{Sk}}$ \cite{polyakov22metastable, buttner2018theory, bernand2022micromagnetic, bernand2023theory}, as well as their perturbative extensions from higher-order expansion terms \cite{Masell2019}.
However, under the constraint of $r_0=3$, as shown in Fig.~\ref{fig_2}(b), $\Delta E$ exhibits a pronounced non-linear dependence on $A$ while $R_\text{Sk}$ remains nearly constant. The non-linear behavior of $\Delta E$ can be attributed to the variation in the spin texture of the SP, which is accompanied by a marked reduction in $R_\text{SP}$.
To further elucidate the deviation from the expected linear scaling, we introduce $J_4$ while keeping $A = 0.5$ meV. This extra degree of freedom leads to an under-determined transformation $(r_0, A)\to(J_2,J_3,J_4)$.
Consequently, as shown in Fig.~\ref{fig_2}(c), a variation of $J_4$ has no influence on either $R_\text{Sk}$ nor $R_\text{SP}$ when $A$ and $r_0$ are kept constant. In contrast, $\Delta E$ exhibits a pronounced variation. 

To explain this, we decompose $\mathbf{J} = \mathbf{J}_{\parallel} + \mathbf{J}_{\perp}$, where $\mathbf{J}_{\parallel}\in\mathrm{span}(\mathbf{a},\mathbf{b})$ sets the micromagnetic parameters, and $\mathbf{J}_{\perp}$ is orthogonal to both $\mathbf{a}$ and $\mathbf{b}$, thus leaving the continuum description unchanged ($\mathbf{c}$ is neglected as $C=4B$). A variation of $J_4$ in Fig.~\ref{fig_2}(c) changes only $\mathbf{J}_{\perp}$, while leaving $\mathbf{J}_{\parallel}$ fixed.
This leads to the conclusion that, while the radius $R_\text{Sk}$ is nearly determined by the parameters $A$ and $r_0$, $\Delta E$ has an additional dependence on $\mathbf{J}_{\perp}$. Thus, for the hexagonal lattice, the statement
\begin{equation}\label{eq:hidden_J}
    R\approx R(r_0,A)~,\quad \Delta E = \Delta E(r_0,A,\mathbf{J}_{\perp})
\end{equation}
consequently implies the existence of degrees of freedom, which can be tuned in order to enhance the energy barrier of a skyrmion of fixed size.

\begin{figure}[b]%---------------------------------------------------------------------------------------------------------------------Figure 3
	\centering
	\includegraphics[width=1.0\linewidth]{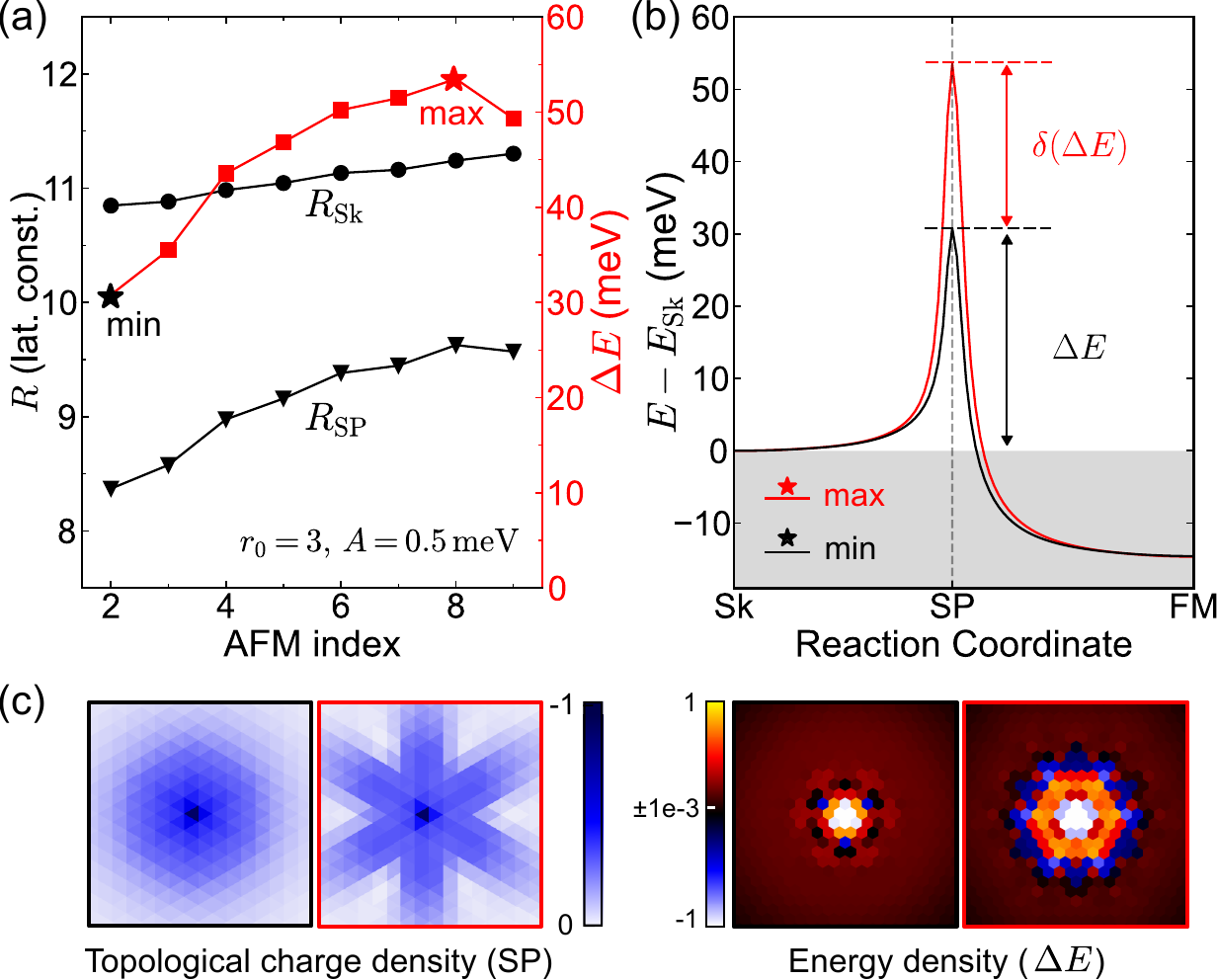}
	\caption{\label{fig_3}
Enhancing skyrmion stability by extending the range of exchange frustration.
(a) Energy barrier $\Delta E$, skyrmion radius $R_\text{Sk}$, and SP radius $R_\text{SP}$ as a function of AFM index. As the AFM term shifts from $J_2$ to $J_9$, $\Delta E$ increases significantly, demonstrating the effect of LR frustration.
(b) MEPs for the minimum (black) and maximum (red) $\Delta E$ cases in (a), showing the increase in $\Delta E$ without changing $E_\text{Sk} - E_\text{FM}$ from SR to LR frustration. (c) Topological charge densities of the SP and energy densities of $\Delta E_{\text{min}}$ and $\Delta E_{\text{max}}$ in (b).
}
\end{figure}

\textit{Long-range frustration.}
From the inset in Fig.~\ref{fig_2}(c), it can be seen that the increasing $\Delta E$ is accompanied by a shift of the negative sign from $J_2$ to $J_3$. 
This suggests that the range of frustration may be a key factor controlling $\Delta E$. In order to examine this, we construct a minimal model with $J_1 = 10$ meV and a single negative constant with variable shell index from $J_2$ to $J_9$ (AFM index) $J_{\uparrow\downarrow}<0$, enabling a controlled transition from SR to LR frustration. A positive $J_{10}$ is introduced to ensure a determined mapping $(r_0, A)\to(J_{\uparrow\downarrow},J_{10})$.
As shown in Fig.~\ref{fig_3}(a), shifting the AFM index to more distant shells strongly enhances $\Delta E$. In particular, the shift from $J_1-J_2$ to $J_1-J_8$ enhances $\Delta E$ from 30 to 53 meV, while at $J_1-J_9$, $\Delta E$ is slightly reduced due to increasing FM contributions from $J_{10}$ (see Sec.~S6 in SM~\cite{supplmat}). 
The corresponding minimum and maximum $\Delta E$, shown in Fig.~\ref{fig_3}(b), further reveal that this transformation leaves the energy with respect to the FM invariant. Since this leaves only modulations of the SP as a possible source of the enhancement, we highlight the respective differences in topological and energy density in Fig.~\ref{fig_3}(c).
The deviations indicate that the increase in $R_\text{SP}$ is not a simple rescaling but rather an atomistic variation. This means that SP's texture and energy can be affected by $\mathbf{J}_{\perp}$, which in turn changes $\Delta E$.

%-----------------------------------------------------------------------------------------------------------------------------

\begin{figure*}[t]%---------------------------------------------------------------------------------------------------------------------Figure 4
	\centering
	\includegraphics[width=1.0\linewidth]{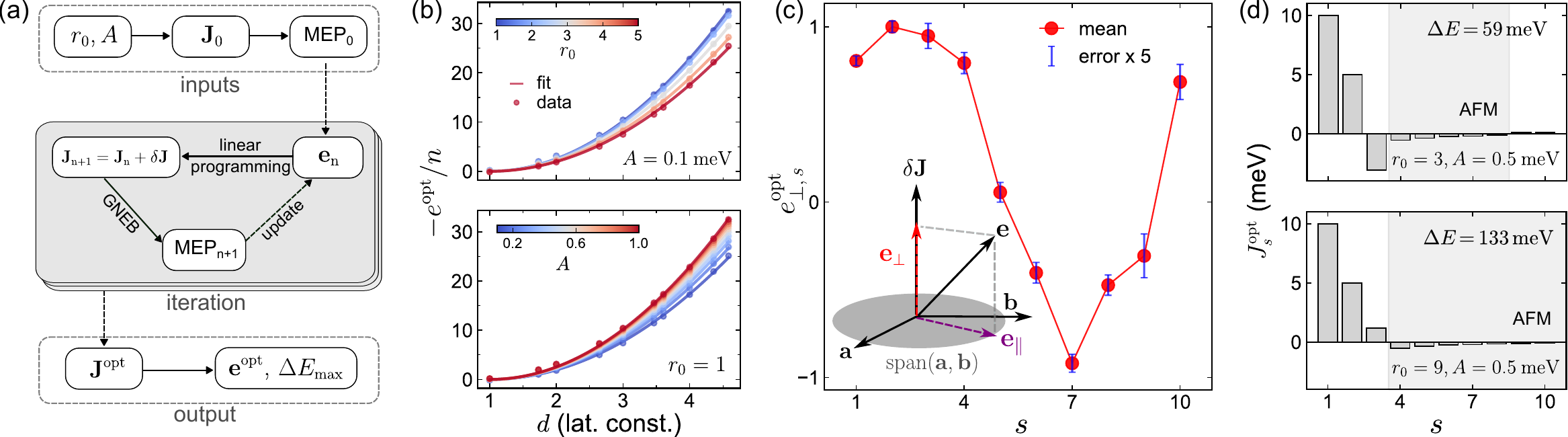}
	\caption{\label{fig_4} 
Energy barrier optimization in exchange space. 
(a) Schematic illustration of the optimization framework. For fixed micromagnetic parameters $(r_0, A)$, the energy barrier is expressed as $\Delta E=\mathbf{e}\cdot\mathbf{J}$, and optimized by iteratively updating $\mathbf{J}$ under physical constraints.
(b) Normalized components of the optimal differential spin-structure vector $-e^\text{opt}/n$ as a function of shell distance $d$ for different micromagnetic parameters. $r_0$ is varied with $A=0.1$ meV fixed (top panel), while in $A$ is varied with $r_0=1$ fixed (bottom panel).
(c) Averaged components of $\mathbf{e}_{\perp}^{\text{opt}}$ for different micromagnetic parameters, rescaled to unit maximum. While $\mathbf{e}$ varies with $(r_0,A)$, $\mathbf{e}_{\perp}$ is nearly invariant, indicating that its direction is mainly determined by lattice geometry.
(d) Optimized exchange parameters for $A=0.5$ meV, $r_0 = 3$ and 9. Although LR contributions are restricted, $\Delta E$ remains significantly enhanced. For a higher $r_0$, the AFM region shifts to the right.
}
\end{figure*}%-----------------------------------------------------------------------------------------------------------------------------

\textit{Energy barrier optimization.}
Building on top of our results on important degrees of freedom, we now formulate an optimization framework to maximize $\Delta E$ by optimizing $\mathbf{J}$.
Eq.~(\ref{eq:exchange_hamiltonian}) allows us to linearize $\Delta E$ as $\Delta E = \mathbf{e} \cdot \mathbf{J}$, where $\mathbf{e}$ is the differential spin-structure vector with components
\begin{equation}
e_s = \sum_{\langle i,j\rangle_s} \Bigl[ (\mathbf{m}_i \cdot \mathbf{m}_j)\big|_{\text{SP}} - (\mathbf{m}_i \cdot \mathbf{m}_j)\big|_{\text{Sk}} \Bigr]\,.
\label{eq:barrier_linear}
\end{equation}
which can be calculated based on the given MEP. As illustrated in Fig.~\ref{fig_4}(a), the optimization starts from an initial guess MEP$_0$ for a given $(r_0,A)$.
To restrict the search to physically realistic interactions, $\mathbf{J}$ is constrained to realistic RKKY-like magnitudes (see SM Sec.~S7 \cite{supplmat}). 
Under these constraints, the feasible set of $\mathbf{J}$ is reduced from an affine subspace to a bounded one, so the optimization reduces to a standard linear programming problem that can be solved using open-source algorithms. A new MEP is then obtained from GNEB and used to update $\mathbf{e}$. The procedure has been solved self-consistently until convergence, yielding the optimized exchange interaction $\mathbf{J}^{\text{opt}}$, $\mathbf{e}^{\text{opt}}$
%\moritz{$\leftarrow$(\textbf{why subscript again, when its superscript everywhere else?})} \shiwei{$\leftarrow$(\textbf{I was using subscript for vectors.})}, 
and the corresponding maximum energy barrier $\Delta E_{\text{max}}$.

We use this framework for skyrmions under micromagnetic parameters with $r_0 \in [1,5]$ and $A \in [0.1,1]$, and analyze the corresponding $\mathbf{e}^{\text{opt}}$. Representative results are shown in Fig.~\ref{fig_4}(b-c). To eliminate the trivial enhancement associated with the coordination number $n_s$, we plot the normalized components $-e_{s}^{\text{opt}}/n_s$. The magnitude of the components rises with shell distance $d$, which we attribute to LR terms capturing more pronounced spin rotations in spin textures. This dependence is well described by $e/n=k(d-1)^2$, where $k$ depends on $(r_0,A)$. 
Increasing $r_0$ reduces $k$ due to the more gradual deflection between neighboring spins for larger skyrmions. In contrast, $k$ increases with $A$ and gradually saturates, accounting for the nonlinear barrier dependence in Fig.~\ref{fig_2}(b). 
The maximum energy barrier can therefore be approximated as
\begin{equation}
    \Delta E_{\text{max}} \approx k\sum_s n_s(d_s-d_1)^2J_s^
    \text{opt},~~ k=k(r_0,A) <0~,
\label{eq:barrier_description}
\end{equation}
with ${e_s^\text{opt}} \approx k n_s(d_s-d_1)^2$. 
Here, $n_s$ and $d_s$ denote the coordination number and shell distance of the $s$-th neighbor shell, respectively, and $d_1$ is the nearest-neighbor distance. 
Details of the parametrization of $k(r_0,A)$ are given in SM Sec.~S8 \cite{supplmat}. 
According to Eq.~(\ref{eq:hidden_J}), the micromagnetic parameters alone are insufficient to uniquely determine $\Delta E$. However, by incorporating the atomic-scale realization $J_s$, a general description of the barrier can still be established. Interestingly, compared with Eq.~(\ref{eq:barrier_linear}), the direction of optimal $\mathbf{e}$ for skyrmions in Eq.~(\ref{eq:barrier_description}) seems determined solely by the lattice information. 
Therefore, in the following, we analyze how the lattice geometry couples to atomic exchange and thereby modifies $\Delta E$.
 
We consider the change to the energy barrier under a small variation $\delta\mathbf{J}$,
\begin{equation}
\delta(\Delta E) = \mathbf{e}\cdot\delta\mathbf{J}
= (\mathbf{e}_{\parallel} + \mathbf{e}_{\perp})\cdot\delta\mathbf{J}~,
\end{equation}
where $\mathbf{e}$ is decomposed into components parallel and perpendicular to $\mathrm{span}(\mathbf{a},\mathbf{b})$.
Variations $\delta\mathbf{J}$ that increase $\Delta E$ while preserving micromagnetic parameters, thus have to fulfill
\begin{equation}
\mathbf{e}_{\parallel}\cdot\delta\mathbf{J}=0~, \quad
\mathbf{e}_{\perp}\cdot\delta\mathbf{J}>0~.
\end{equation}
Here, $\mathbf{e}_{\perp}$ defines the optimal direction in exchange space that maximizes $\Delta E$ while keeping skyrmion size and magnetic energy scale unchanged.
We compute $\mathbf{e}_{\perp}^{\text{opt}}$
%\moritz{\textbf{(shouldn't it be $\mathbf{e}_{\perp}^{\text{opt}}$, because this is what is shown in Fig.4c, right?)}} 
for skyrmions under different micromagnetic parameters and rescale them to unit maximum. Fig.~\ref{fig_4}(c) shows the corresponding averages and error bars.
While $\mathbf{e}$ varies with micromagnetic parameters, $\mathbf{e}_{\perp}$ is nearly unchanged, since the factor $k(r_0,A)$ in Eq.~\eqref{eq:barrier_description} does not affect its direction. Positive values for $s\leq5$ suggest that strengthening FM exchange in those shells enhances $\Delta E$, whereas negative values suggest stronger AFM exchange for $J_6-J_9$. 

We further compare the influence of $\delta\mathbf{J}$ on both the skyrmion and SP energies to clarify the origin of $\mathbf{e}_\perp$ (see SM Sec.~S9~\cite{supplmat}). Since we find that it affects only the SP energy and leaves the skyrmion energy unchanged, our method enables the construction of $\delta\mathbf{J}$ with arbitrary magnitude to enhance $\Delta E$ while preserving the metastable skyrmion.
We also calculated $\mathbf{e}_{\perp}$ for both square and honeycomb lattices in SM Sec.~S10 \cite{supplmat}. 
In both cases, the negative components of $\mathbf{e}_{\perp}$ consistently reside in the $J_6$--$J_9$ shells. This indicates that LR frustration is a generic and effective mechanism for enhancing skyrmion stability across different lattice geometries.

However, due to physical constraints, $\mathbf{J}^{\text{opt}}$ cannot fully align with the direction of $\mathbf{e}_{\perp}$. The optimized exchange interactions for $A=0.5$ meV, $r_0 = 3$ and 9 are shown in Fig.~\ref{fig_4}(d). Even weak LR terms introduced on top of SR frustration can enhance $\Delta E$ by approximately a factor of two compared to the SR case [see Fig.~\ref{fig_3}(a)]. Although the direction of $\mathbf{e}_{\perp}$ is nearly independent of $r_0$, the optimized interactions exhibit a progressive shift of the AFM weight toward more distant shells with increasing $r_0$, indicating an enhanced role of LR frustration for larger skyrmions. As discussed in SM Sec.~S11 \cite{supplmat}, this behavior originates from $\mathbf{J}_{\parallel}$, which is uniquely determined by the micromagnetic constraints and naturally favors LR frustration.

% % %========================================================
% % % conclusion
% % %========================================================

In summary, we establish LR frustration as a route beyond conventional micromagnetic limits, yielding strongly enhanced skyrmion stability at fixed size and energy scale.
From a materials perspective, LR exchange interactions are naturally present in a wide range of systems, including ultrathin films and 2D van der Waals magnets~\cite{von2017enhanced, goerzen2023lifetime, Dongzhe2022_fgt, Moritz_prb2026}.
Although often treated as weak corrections, our results highlight their decisive role in skyrmion stability, consistent with recent studies showing stabilization of diverse topological spin textures by LR frustration as small as 1–2\% of $J_1$~\cite{shiwei2026}.
We anticipate that our work will motivate experimental studies of LR frustration, for instance, through combined spin-wave spectroscopy and skyrmion lifetime measurements. As LR exchange interactions can be tuned via interface engineering, strain, and stacking design \cite{zhang20242d}, our results identify LR exchange engineering as a practical route toward robust nanoscale skyrmions for device applications.

% 	%%%%%%%%%%%%%%%%% Acknowledgement %%%%%%%%%%%%%%%%%%%%%%%

\section*{ACKNOWLEDGMENTS}
This work is supported by the National Natural Science Foundation of China (Grant No.11804301), the Natural Science Foundation of Zhejiang Province (Grant No. LMS25A040001), the Funds of the Natural Science Foundation of Hangzhou (Grant No. 2025SZRJJ0830), and France 2030 government investment plan managed by the French National Research Agency under grant reference PEPR SPIN – [SPINTHEORY] ANR-22-EXSP-0009. This study has been (partially) supported through the grant NanoX no.~ANR-17-EURE-0009 in the framework of the ``Programme des Investissements d’Avenir". This work was performed using HPC resources from CALMIP (Grant No. 2024/2026-[P21008]). We thank F. N. Rybakov and N. S. Kiselev for valuable discussions.

\bibliography{References}

\begin{thebibliography}{48}%
\makeatletter
\providecommand \@ifxundefined [1]{%
 \@ifx{#1\undefined}
}%
\providecommand \@ifnum [1]{%
 \ifnum #1\expandafter \@firstoftwo
 \else \expandafter \@secondoftwo
 \fi
}%
\providecommand \@ifx [1]{%
 \ifx #1\expandafter \@firstoftwo
 \else \expandafter \@secondoftwo
 \fi
}%
\providecommand \natexlab [1]{#1}%
\providecommand \enquote  [1]{``#1''}%
\providecommand \bibnamefont  [1]{#1}%
\providecommand \bibfnamefont [1]{#1}%
\providecommand \citenamefont [1]{#1}%
\providecommand \href@noop [0]{\@secondoftwo}%
\providecommand \href [0]{\begingroup \@sanitize@url \@href}%
\providecommand \@href[1]{\@@startlink{#1}\@@href}%
\providecommand \@@href[1]{\endgroup#1\@@endlink}%
\providecommand \@sanitize@url [0]{\catcode `\\12\catcode `\$12\catcode
  `\&12\catcode `\#12\catcode `\^12\catcode `\_12\catcode `\%12\relax}%
\providecommand \@@startlink[1]{}%
\providecommand \@@endlink[0]{}%
\providecommand \url  [0]{\begingroup\@sanitize@url \@url }%
\providecommand \@url [1]{\endgroup\@href {#1}{\urlprefix }}%
\providecommand \urlprefix  [0]{URL }%
\providecommand \Eprint [0]{\href }%
\providecommand \doibase [0]{https://doi.org/}%
\providecommand \selectlanguage [0]{\@gobble}%
\providecommand \bibinfo  [0]{\@secondoftwo}%
\providecommand \bibfield  [0]{\@secondoftwo}%
\providecommand \translation [1]{[#1]}%
\providecommand \BibitemOpen [0]{}%
\providecommand \bibitemStop [0]{}%
\providecommand \bibitemNoStop [0]{.\EOS\space}%
\providecommand \EOS [0]{\spacefactor3000\relax}%
\providecommand \BibitemShut  [1]{\csname bibitem#1\endcsname}%
\let\auto@bib@innerbib\@empty
%</preamble>
\bibitem [{\citenamefont {Heinze}\ \emph {et~al.}(2011)\citenamefont {Heinze},
  \citenamefont {Von~Bergmann}, \citenamefont {Menzel}, \citenamefont {Brede},
  \citenamefont {Kubetzka}, \citenamefont {Wiesendanger}, \citenamefont
  {Bihlmayer},\ and\ \citenamefont {Bl{\"u}gel}}]{heinze2011spontaneous}%
  \BibitemOpen
  \bibfield  {author} {\bibinfo {author} {\bibfnamefont {S.}~\bibnamefont
  {Heinze}}, \bibinfo {author} {\bibfnamefont {K.}~\bibnamefont
  {Von~Bergmann}}, \bibinfo {author} {\bibfnamefont {M.}~\bibnamefont
  {Menzel}}, \bibinfo {author} {\bibfnamefont {J.}~\bibnamefont {Brede}},
  \bibinfo {author} {\bibfnamefont {A.}~\bibnamefont {Kubetzka}}, \bibinfo
  {author} {\bibfnamefont {R.}~\bibnamefont {Wiesendanger}}, \bibinfo {author}
  {\bibfnamefont {G.}~\bibnamefont {Bihlmayer}},\ and\ \bibinfo {author}
  {\bibfnamefont {S.}~\bibnamefont {Bl{\"u}gel}},\ }\bibfield  {title}
  {\bibinfo {title} {Spontaneous atomic-scale magnetic skyrmion lattice in two
  dimensions},\ }\href {https://doi.org/10.1038/NPHYS2045} {\bibfield
  {journal} {\bibinfo  {journal} {Nat. Phys.}\ }\textbf {\bibinfo {volume}
  {7}},\ \bibinfo {pages} {713} (\bibinfo {year} {2011})}\BibitemShut {NoStop}%
\bibitem [{\citenamefont {Romming}\ \emph {et~al.}(2013)\citenamefont
  {Romming}, \citenamefont {Hanneken}, \citenamefont {Menzel}, \citenamefont
  {Bickel}, \citenamefont {Wolter}, \citenamefont {von Bergmann}, \citenamefont
  {Kubetzka},\ and\ \citenamefont {Wiesendanger}}]{Romming2013}%
  \BibitemOpen
  \bibfield  {author} {\bibinfo {author} {\bibfnamefont {N.}~\bibnamefont
  {Romming}}, \bibinfo {author} {\bibfnamefont {C.}~\bibnamefont {Hanneken}},
  \bibinfo {author} {\bibfnamefont {M.}~\bibnamefont {Menzel}}, \bibinfo
  {author} {\bibfnamefont {J.}~\bibnamefont {Bickel}}, \bibinfo {author}
  {\bibfnamefont {B.}~\bibnamefont {Wolter}}, \bibinfo {author} {\bibfnamefont
  {K.}~\bibnamefont {von Bergmann}}, \bibinfo {author} {\bibfnamefont
  {A.}~\bibnamefont {Kubetzka}},\ and\ \bibinfo {author} {\bibfnamefont
  {R.}~\bibnamefont {Wiesendanger}},\ }\bibfield  {title} {\bibinfo {title}
  {Writing and deleting single magnetic skyrmions},\ }\href
  {https://doi.org/10.1126/science.1240573} {\bibfield  {journal} {\bibinfo
  {journal} {Science}\ }\textbf {\bibinfo {volume} {341}},\ \bibinfo {pages}
  {636} (\bibinfo {year} {2013})}\BibitemShut {NoStop}%
\bibitem [{\citenamefont {Li}\ \emph {et~al.}(2022)\citenamefont {Li},
  \citenamefont {Haldar},\ and\ \citenamefont {Heinze}}]{Dongzhe2022_fgt}%
  \BibitemOpen
  \bibfield  {author} {\bibinfo {author} {\bibfnamefont {D.}~\bibnamefont
  {Li}}, \bibinfo {author} {\bibfnamefont {S.}~\bibnamefont {Haldar}},\ and\
  \bibinfo {author} {\bibfnamefont {S.}~\bibnamefont {Heinze}},\ }\bibfield
  {title} {\bibinfo {title} {Strain-driven zero-field near-10 nm skyrmions in
  two-dimensional van der {Waals} heterostructures},\ }\href
  {https://doi.org/10.1021/acs.nanolett.2c03287} {\bibfield  {journal}
  {\bibinfo  {journal} {Nano Lett.}\ }\textbf {\bibinfo {volume} {22}},\
  \bibinfo {pages} {7706} (\bibinfo {year} {2022})}\BibitemShut {NoStop}%
\bibitem [{\citenamefont {Fert}\ \emph
  {et~al.}(2013{\natexlab{a}})\citenamefont {Fert}, \citenamefont {Cros},\ and\
  \citenamefont {Sampaio}}]{fert2013}%
  \BibitemOpen
  \bibfield  {author} {\bibinfo {author} {\bibfnamefont {A.}~\bibnamefont
  {Fert}}, \bibinfo {author} {\bibfnamefont {V.}~\bibnamefont {Cros}},\ and\
  \bibinfo {author} {\bibfnamefont {J.}~\bibnamefont {Sampaio}},\ }\bibfield
  {title} {\bibinfo {title} {{Skyrmions on the track}},\ }\href
  {https://doi.org/10.1038/nnano.2013.29} {\bibfield  {journal} {\bibinfo
  {journal} {Nat. Nanotechnol.}\ }\textbf {\bibinfo {volume} {8}},\ \bibinfo
  {pages} {152} (\bibinfo {year} {2013}{\natexlab{a}})}\BibitemShut {NoStop}%
\bibitem [{\citenamefont {Sampaio}\ \emph {et~al.}(2013)\citenamefont
  {Sampaio}, \citenamefont {Cros}, \citenamefont {Rohart}, \citenamefont
  {Thiaville},\ and\ \citenamefont {Fert}}]{sampaio2013nucleation}%
  \BibitemOpen
  \bibfield  {author} {\bibinfo {author} {\bibfnamefont {J.}~\bibnamefont
  {Sampaio}}, \bibinfo {author} {\bibfnamefont {V.}~\bibnamefont {Cros}},
  \bibinfo {author} {\bibfnamefont {S.}~\bibnamefont {Rohart}}, \bibinfo
  {author} {\bibfnamefont {A.}~\bibnamefont {Thiaville}},\ and\ \bibinfo
  {author} {\bibfnamefont {A.}~\bibnamefont {Fert}},\ }\bibfield  {title}
  {\bibinfo {title} {Nucleation, stability and current-induced motion of
  isolated magnetic skyrmions in nanostructures},\ }\href
  {https://doi.org/10.1038/NNANO.2013.210} {\bibfield  {journal} {\bibinfo
  {journal} {Nat. Nanotechnol.}\ }\textbf {\bibinfo {volume} {8}},\ \bibinfo
  {pages} {839} (\bibinfo {year} {2013})}\BibitemShut {NoStop}%
\bibitem [{\citenamefont {Larra{\~n}aga}\ \emph {et~al.}(2024)\citenamefont
  {Larra{\~n}aga} \emph {et~al.}}]{urrestarazu2024electrical}%
  \BibitemOpen
  \bibfield  {author} {\bibinfo {author} {\bibfnamefont {U.}~\bibnamefont
  {Larra{\~n}aga}} \emph {et~al.},\ }\bibfield  {title} {\bibinfo {title}
  {Electrical detection and nucleation of a magnetic skyrmion in a magnetic
  tunnel junction observed via operando magnetic microscopy},\ }\href
  {https://doi.org/10.1021/acs.nanolett.4c00316} {\bibfield  {journal}
  {\bibinfo  {journal} {Nano Lett.}\ }\textbf {\bibinfo {volume} {24}},\
  \bibinfo {pages} {3557} (\bibinfo {year} {2024})}\BibitemShut {NoStop}%
\bibitem [{\citenamefont {Li}\ \emph {et~al.}(2024)\citenamefont {Li},
  \citenamefont {Haldar},\ and\ \citenamefont {Heinze}}]{li2024proposal}%
  \BibitemOpen
  \bibfield  {author} {\bibinfo {author} {\bibfnamefont {D.}~\bibnamefont
  {Li}}, \bibinfo {author} {\bibfnamefont {S.}~\bibnamefont {Haldar}},\ and\
  \bibinfo {author} {\bibfnamefont {S.}~\bibnamefont {Heinze}},\ }\bibfield
  {title} {\bibinfo {title} {Proposal for all-electrical skyrmion detection in
  van der {Waals} tunnel junctions},\ }\href
  {https://doi.org/10.1021/acs.nanolett.3c04238} {\bibfield  {journal}
  {\bibinfo  {journal} {Nano Lett.}\ }\textbf {\bibinfo {volume} {24}},\
  \bibinfo {pages} {2496} (\bibinfo {year} {2024})}\BibitemShut {NoStop}%
\bibitem [{\citenamefont {Chen}\ \emph {et~al.}(2024)\citenamefont {Chen} \emph
  {et~al.}}]{chen2024all}%
  \BibitemOpen
  \bibfield  {author} {\bibinfo {author} {\bibfnamefont {S.}~\bibnamefont
  {Chen}} \emph {et~al.},\ }\bibfield  {title} {\bibinfo {title}
  {All-electrical skyrmionic magnetic tunnel junction},\ }\href
  {https://doi.org/10.1038/s41586-024-07131-7} {\bibfield  {journal} {\bibinfo
  {journal} {Nature}\ }\textbf {\bibinfo {volume} {627}},\ \bibinfo {pages}
  {522} (\bibinfo {year} {2024})}\BibitemShut {NoStop}%
\bibitem [{\citenamefont {Buttner}\ \emph {et~al.}(2021)\citenamefont
  {Buttner}, \citenamefont {Pfau}, \citenamefont {Bottcher}, \citenamefont
  {Schneider}, \citenamefont {Mercurio}, \citenamefont {Gunther}, \citenamefont
  {Hessing}, \citenamefont {Klose}, \citenamefont {Wittmann}, \citenamefont
  {Gerlinger} \emph {et~al.}}]{buttner2021observation}%
  \BibitemOpen
  \bibfield  {author} {\bibinfo {author} {\bibfnamefont {F.}~\bibnamefont
  {Buttner}}, \bibinfo {author} {\bibfnamefont {B.}~\bibnamefont {Pfau}},
  \bibinfo {author} {\bibfnamefont {M.}~\bibnamefont {Bottcher}}, \bibinfo
  {author} {\bibfnamefont {M.}~\bibnamefont {Schneider}}, \bibinfo {author}
  {\bibfnamefont {G.}~\bibnamefont {Mercurio}}, \bibinfo {author}
  {\bibfnamefont {C.~M.}\ \bibnamefont {Gunther}}, \bibinfo {author}
  {\bibfnamefont {P.}~\bibnamefont {Hessing}}, \bibinfo {author} {\bibfnamefont
  {C.}~\bibnamefont {Klose}}, \bibinfo {author} {\bibfnamefont
  {A.}~\bibnamefont {Wittmann}}, \bibinfo {author} {\bibfnamefont
  {K.}~\bibnamefont {Gerlinger}}, \emph {et~al.},\ }\bibfield  {title}
  {\bibinfo {title} {Observation of fluctuation-mediated picosecond nucleation
  of a topological phase},\ }\href {https://doi.org/10.1038/s41563-020-00807-1}
  {\bibfield  {journal} {\bibinfo  {journal} {Nat. Mater.}\ }\textbf {\bibinfo
  {volume} {20}},\ \bibinfo {pages} {30} (\bibinfo {year} {2021})}\BibitemShut
  {NoStop}%
\bibitem [{\citenamefont {Dabrowski}\ \emph {et~al.}(2022)\citenamefont
  {Dabrowski}, \citenamefont {Guo}, \citenamefont {Strungaru}, \citenamefont
  {Keatley}, \citenamefont {Withers}, \citenamefont {Santos},\ and\
  \citenamefont {Hicken}}]{dabrowski2022all}%
  \BibitemOpen
  \bibfield  {author} {\bibinfo {author} {\bibfnamefont {M.}~\bibnamefont
  {Dabrowski}}, \bibinfo {author} {\bibfnamefont {S.}~\bibnamefont {Guo}},
  \bibinfo {author} {\bibfnamefont {M.}~\bibnamefont {Strungaru}}, \bibinfo
  {author} {\bibfnamefont {P.~S.}\ \bibnamefont {Keatley}}, \bibinfo {author}
  {\bibfnamefont {F.}~\bibnamefont {Withers}}, \bibinfo {author} {\bibfnamefont
  {E.~J.}\ \bibnamefont {Santos}},\ and\ \bibinfo {author} {\bibfnamefont
  {R.~J.}\ \bibnamefont {Hicken}},\ }\bibfield  {title} {\bibinfo {title}
  {All-optical control of spin in a {2D} van der {W}aals magnet},\ }\href
  {https://doi.org/10.1038/s41467-022-33343-4} {\bibfield  {journal} {\bibinfo
  {journal} {Nat. Commun.}\ }\textbf {\bibinfo {volume} {13}},\ \bibinfo
  {pages} {5976} (\bibinfo {year} {2022})}\BibitemShut {NoStop}%
\bibitem [{\citenamefont {Khela}\ \emph {et~al.}(2023)\citenamefont {Khela},
  \citenamefont {Dabrowski}, \citenamefont {Khan}, \citenamefont {Keatley},
  \citenamefont {Verzhbitskiy}, \citenamefont {Eda}, \citenamefont {Hicken},
  \citenamefont {Kurebayashi},\ and\ \citenamefont {Santos}}]{khela2023laser}%
  \BibitemOpen
  \bibfield  {author} {\bibinfo {author} {\bibfnamefont {M.}~\bibnamefont
  {Khela}}, \bibinfo {author} {\bibfnamefont {M.}~\bibnamefont {Dabrowski}},
  \bibinfo {author} {\bibfnamefont {S.}~\bibnamefont {Khan}}, \bibinfo {author}
  {\bibfnamefont {P.~S.}\ \bibnamefont {Keatley}}, \bibinfo {author}
  {\bibfnamefont {I.}~\bibnamefont {Verzhbitskiy}}, \bibinfo {author}
  {\bibfnamefont {G.}~\bibnamefont {Eda}}, \bibinfo {author} {\bibfnamefont
  {R.~J.}\ \bibnamefont {Hicken}}, \bibinfo {author} {\bibfnamefont
  {H.}~\bibnamefont {Kurebayashi}},\ and\ \bibinfo {author} {\bibfnamefont
  {E.~J.}\ \bibnamefont {Santos}},\ }\bibfield  {title} {\bibinfo {title}
  {Laser-induced topological spin switching in a {2D} van der {W}aals magnet},\
  }\href {https://doi.org/10.1038/s41467-023-37082-y} {\bibfield  {journal}
  {\bibinfo  {journal} {Nat. Commun.}\ }\textbf {\bibinfo {volume} {14}},\
  \bibinfo {pages} {1378} (\bibinfo {year} {2023})}\BibitemShut {NoStop}%
\bibitem [{\citenamefont {Fert}\ \emph
  {et~al.}(2013{\natexlab{b}})\citenamefont {Fert}, \citenamefont {Cros},\ and\
  \citenamefont {Sampaio}}]{fert2013skyrmions}%
  \BibitemOpen
  \bibfield  {author} {\bibinfo {author} {\bibfnamefont {A.}~\bibnamefont
  {Fert}}, \bibinfo {author} {\bibfnamefont {V.}~\bibnamefont {Cros}},\ and\
  \bibinfo {author} {\bibfnamefont {J.}~\bibnamefont {Sampaio}},\ }\bibfield
  {title} {\bibinfo {title} {Skyrmions on the track},\ }\href
  {https://doi.org/10.1038/nnano.2013.29} {\bibfield  {journal} {\bibinfo
  {journal} {Nat. Nanotechnol.}\ }\textbf {\bibinfo {volume} {8}},\ \bibinfo
  {pages} {152} (\bibinfo {year} {2013}{\natexlab{b}})}\BibitemShut {NoStop}%
\bibitem [{\citenamefont {Fert}\ \emph {et~al.}(2017)\citenamefont {Fert},
  \citenamefont {Reyren},\ and\ \citenamefont {Cros}}]{fert2017magnetic}%
  \BibitemOpen
  \bibfield  {author} {\bibinfo {author} {\bibfnamefont {A.}~\bibnamefont
  {Fert}}, \bibinfo {author} {\bibfnamefont {N.}~\bibnamefont {Reyren}},\ and\
  \bibinfo {author} {\bibfnamefont {V.}~\bibnamefont {Cros}},\ }\bibfield
  {title} {\bibinfo {title} {Magnetic skyrmions: advances in physics and
  potential applications},\ }\href {https://doi.org/10.1038/natrevmats.2017.31}
  {\bibfield  {journal} {\bibinfo  {journal} {Nat. Rev. Mater.}\ }\textbf
  {\bibinfo {volume} {2}},\ \bibinfo {pages} {1} (\bibinfo {year}
  {2017})}\BibitemShut {NoStop}%
\bibitem [{\citenamefont {Back}\ \emph {et~al.}(2020)\citenamefont {Back},
  \citenamefont {Cros}, \citenamefont {Ebert}, \citenamefont {Everschor-Sitte},
  \citenamefont {Fert}, \citenamefont {Garst}, \citenamefont {Ma},
  \citenamefont {Mankovsky}, \citenamefont {Monchesky}, \citenamefont
  {Mostovoy} \emph {et~al.}}]{back20202020}%
  \BibitemOpen
  \bibfield  {author} {\bibinfo {author} {\bibfnamefont {C.}~\bibnamefont
  {Back}}, \bibinfo {author} {\bibfnamefont {V.}~\bibnamefont {Cros}}, \bibinfo
  {author} {\bibfnamefont {H.}~\bibnamefont {Ebert}}, \bibinfo {author}
  {\bibfnamefont {K.}~\bibnamefont {Everschor-Sitte}}, \bibinfo {author}
  {\bibfnamefont {A.}~\bibnamefont {Fert}}, \bibinfo {author} {\bibfnamefont
  {M.}~\bibnamefont {Garst}}, \bibinfo {author} {\bibfnamefont
  {T.}~\bibnamefont {Ma}}, \bibinfo {author} {\bibfnamefont {S.}~\bibnamefont
  {Mankovsky}}, \bibinfo {author} {\bibfnamefont {T.}~\bibnamefont
  {Monchesky}}, \bibinfo {author} {\bibfnamefont {M.}~\bibnamefont {Mostovoy}},
  \emph {et~al.},\ }\bibfield  {title} {\bibinfo {title} {The 2020 skyrmionics
  roadmap},\ }\href {https://doi.org/10.1088/1361-6463/ab8418} {\bibfield
  {journal} {\bibinfo  {journal} {J. Phys. D:Appl. Phys.}\ }\textbf {\bibinfo
  {volume} {53}},\ \bibinfo {pages} {363001} (\bibinfo {year}
  {2020})}\BibitemShut {NoStop}%
\bibitem [{\citenamefont {G{\"o}bel}\ \emph {et~al.}(2021)\citenamefont
  {G{\"o}bel}, \citenamefont {Mertig},\ and\ \citenamefont
  {Tretiakov}}]{gobel2021beyond}%
  \BibitemOpen
  \bibfield  {author} {\bibinfo {author} {\bibfnamefont {B.}~\bibnamefont
  {G{\"o}bel}}, \bibinfo {author} {\bibfnamefont {I.}~\bibnamefont {Mertig}},\
  and\ \bibinfo {author} {\bibfnamefont {O.~A.}\ \bibnamefont {Tretiakov}},\
  }\bibfield  {title} {\bibinfo {title} {Beyond skyrmions: Review and
  perspectives of alternative magnetic quasiparticles},\ }\href
  {https://doi.org/10.1016/j.physrep.2020.10.001} {\bibfield  {journal}
  {\bibinfo  {journal} {Phys. Rep.}\ }\textbf {\bibinfo {volume} {895}},\
  \bibinfo {pages} {1} (\bibinfo {year} {2021})}\BibitemShut {NoStop}%
\bibitem [{\citenamefont {Bernand-Mantel}\ \emph {et~al.}(2020)\citenamefont
  {Bernand-Mantel}, \citenamefont {Muratov},\ and\ \citenamefont
  {Simon}}]{PhysRevB.101.045416}%
  \BibitemOpen
  \bibfield  {author} {\bibinfo {author} {\bibfnamefont {A.}~\bibnamefont
  {Bernand-Mantel}}, \bibinfo {author} {\bibfnamefont {C.~B.}\ \bibnamefont
  {Muratov}},\ and\ \bibinfo {author} {\bibfnamefont {T.~M.}\ \bibnamefont
  {Simon}},\ }\bibfield  {title} {\bibinfo {title} {Unraveling the role of
  dipolar versus dzyaloshinskii-moriya interactions in stabilizing compact
  magnetic skyrmions},\ }\href {https://doi.org/10.1103/PhysRevB.101.045416}
  {\bibfield  {journal} {\bibinfo  {journal} {Phys. Rev. B}\ }\textbf {\bibinfo
  {volume} {101}},\ \bibinfo {pages} {045416} (\bibinfo {year}
  {2020})}\BibitemShut {NoStop}%
\bibitem [{\citenamefont {Paul}\ \emph {et~al.}(2020)\citenamefont {Paul},
  \citenamefont {Haldar}, \citenamefont {von Malottki},\ and\ \citenamefont
  {Heinze}}]{paul2020role}%
  \BibitemOpen
  \bibfield  {author} {\bibinfo {author} {\bibfnamefont {S.}~\bibnamefont
  {Paul}}, \bibinfo {author} {\bibfnamefont {S.}~\bibnamefont {Haldar}},
  \bibinfo {author} {\bibfnamefont {S.}~\bibnamefont {von Malottki}},\ and\
  \bibinfo {author} {\bibfnamefont {S.}~\bibnamefont {Heinze}},\ }\bibfield
  {title} {\bibinfo {title} {Role of higher-order exchange interactions for
  skyrmion stability},\ }\href
  {https://doi.org/https://doi.org/10.1038/s41467-020-18473-x} {\bibfield
  {journal} {\bibinfo  {journal} {Nat. Commun.}\ }\textbf {\bibinfo {volume}
  {11}},\ \bibinfo {pages} {4756} (\bibinfo {year} {2020})}\BibitemShut
  {NoStop}%
\bibitem [{\citenamefont {von Malottki}\ \emph {et~al.}(2017)\citenamefont {von
  Malottki}, \citenamefont {Dup{\'e}}, \citenamefont {Bessarab}, \citenamefont
  {Delin},\ and\ \citenamefont {Heinze}}]{von2017enhanced}%
  \BibitemOpen
  \bibfield  {author} {\bibinfo {author} {\bibfnamefont {S.}~\bibnamefont {von
  Malottki}}, \bibinfo {author} {\bibfnamefont {B.}~\bibnamefont {Dup{\'e}}},
  \bibinfo {author} {\bibfnamefont {P.~F.}\ \bibnamefont {Bessarab}}, \bibinfo
  {author} {\bibfnamefont {A.}~\bibnamefont {Delin}},\ and\ \bibinfo {author}
  {\bibfnamefont {S.}~\bibnamefont {Heinze}},\ }\bibfield  {title} {\bibinfo
  {title} {Enhanced skyrmion stability due to exchange frustration},\ }\href
  {https://doi.org/10.1038/s41598-017-12525-x} {\bibfield  {journal} {\bibinfo
  {journal} {Sci. Rep.}\ }\textbf {\bibinfo {volume} {7}},\ \bibinfo {pages}
  {12299} (\bibinfo {year} {2017})}\BibitemShut {NoStop}%
\bibitem [{\citenamefont {Varentcova}\ \emph {et~al.}(2020)\citenamefont
  {Varentcova}, \citenamefont {von Malottki}, \citenamefont {Potkina},
  \citenamefont {Kwiatkowski}, \citenamefont {Heinze},\ and\ \citenamefont
  {Bessarab}}]{varentcova2020toward}%
  \BibitemOpen
  \bibfield  {author} {\bibinfo {author} {\bibfnamefont {A.~S.}\ \bibnamefont
  {Varentcova}}, \bibinfo {author} {\bibfnamefont {S.}~\bibnamefont {von
  Malottki}}, \bibinfo {author} {\bibfnamefont {M.~N.}\ \bibnamefont
  {Potkina}}, \bibinfo {author} {\bibfnamefont {G.}~\bibnamefont
  {Kwiatkowski}}, \bibinfo {author} {\bibfnamefont {S.}~\bibnamefont
  {Heinze}},\ and\ \bibinfo {author} {\bibfnamefont {P.~F.}\ \bibnamefont
  {Bessarab}},\ }\bibfield  {title} {\bibinfo {title} {Toward room-temperature
  nanoscale skyrmions in ultrathin films},\ }\href
  {https://doi.org/https://doi.org/10.1038/s41524-020-00453-w} {\bibfield
  {journal} {\bibinfo  {journal} {npj Comput. Mater.}\ }\textbf {\bibinfo
  {volume} {6}},\ \bibinfo {pages} {193} (\bibinfo {year} {2020})}\BibitemShut
  {NoStop}%
\bibitem [{\citenamefont {Varentcova}\ \emph {et~al.}(2018)\citenamefont
  {Varentcova}, \citenamefont {Potkina}, \citenamefont {Malottki},
  \citenamefont {Heinze},\ and\ \citenamefont
  {Bessarab}}]{varentsova2018interplay}%
  \BibitemOpen
  \bibfield  {author} {\bibinfo {author} {\bibfnamefont {A.}~\bibnamefont
  {Varentcova}}, \bibinfo {author} {\bibfnamefont {M.}~\bibnamefont {Potkina}},
  \bibinfo {author} {\bibfnamefont {S.}~\bibnamefont {Malottki}}, \bibinfo
  {author} {\bibfnamefont {S.}~\bibnamefont {Heinze}},\ and\ \bibinfo {author}
  {\bibfnamefont {P.}~\bibnamefont {Bessarab}},\ }\bibfield  {title} {\bibinfo
  {title} {Interplay between size and stability of magnetic skyrmions},\ }\href
  {https://doi.org/10.17586/2220-8054-2018-9-3-356-363} {\bibfield  {journal}
  {\bibinfo  {journal} {Nanosyst.:Phys., Chem., Math.}\ }\textbf {\bibinfo
  {volume} {9}},\ \bibinfo {pages} {356} (\bibinfo {year} {2018})}\BibitemShut
  {NoStop}%
\bibitem [{\citenamefont {Haldar}\ \emph {et~al.}(2018)\citenamefont {Haldar},
  \citenamefont {von Malottki}, \citenamefont {Meyer}, \citenamefont
  {Bessarab},\ and\ \citenamefont {Heinze}}]{Soum_2018}%
  \BibitemOpen
  \bibfield  {author} {\bibinfo {author} {\bibfnamefont {S.}~\bibnamefont
  {Haldar}}, \bibinfo {author} {\bibfnamefont {S.}~\bibnamefont {von
  Malottki}}, \bibinfo {author} {\bibfnamefont {S.}~\bibnamefont {Meyer}},
  \bibinfo {author} {\bibfnamefont {P.~F.}\ \bibnamefont {Bessarab}},\ and\
  \bibinfo {author} {\bibfnamefont {S.}~\bibnamefont {Heinze}},\ }\bibfield
  {title} {\bibinfo {title} {{First-principles prediction of sub-10-nm
  skyrmions in Pd/Fe bilayers on Rh(111)}},\ }\href
  {https://doi.org/10.1103/PhysRevB.98.060413} {\bibfield  {journal} {\bibinfo
  {journal} {Phys. Rev. B}\ }\textbf {\bibinfo {volume} {98}},\ \bibinfo
  {pages} {060413} (\bibinfo {year} {2018})}\BibitemShut {NoStop}%
\bibitem [{\citenamefont {Arya}\ \emph {et~al.}(2025)\citenamefont {Arya},
  \citenamefont {Goerzen}, \citenamefont {Calmels}, \citenamefont {Arras},
  \citenamefont {Haldar}, \citenamefont {Heinze},\ and\ \citenamefont
  {Li}}]{megha2025}%
  \BibitemOpen
  \bibfield  {author} {\bibinfo {author} {\bibfnamefont {M.}~\bibnamefont
  {Arya}}, \bibinfo {author} {\bibfnamefont {M.~A.}\ \bibnamefont {Goerzen}},
  \bibinfo {author} {\bibfnamefont {L.}~\bibnamefont {Calmels}}, \bibinfo
  {author} {\bibfnamefont {R.}~\bibnamefont {Arras}}, \bibinfo {author}
  {\bibfnamefont {S.}~\bibnamefont {Haldar}}, \bibinfo {author} {\bibfnamefont
  {S.}~\bibnamefont {Heinze}},\ and\ \bibinfo {author} {\bibfnamefont
  {D.}~\bibnamefont {Li}},\ }\bibfield  {title} {\bibinfo {title} {{A new
  skyrmion topological transition driven by higher-order exchange interactions
  in Janus MnSeTe}},\ }\href {https://doi.org/10.1021/acs.nanolett.5c04452}
  {\bibfield  {journal} {\bibinfo  {journal} {Nano Lett.}\ }\textbf {\bibinfo
  {volume} {25}},\ \bibinfo {pages} {16703} (\bibinfo {year}
  {2025})}\BibitemShut {NoStop}%
\bibitem [{\citenamefont {Haldar}\ \emph {et~al.}(2026)\citenamefont {Haldar},
  \citenamefont {Goerzen}, \citenamefont {Heinze},\ and\ \citenamefont
  {Li}}]{Soum2026}%
  \BibitemOpen
  \bibfield  {author} {\bibinfo {author} {\bibfnamefont {S.}~\bibnamefont
  {Haldar}}, \bibinfo {author} {\bibfnamefont {M.~A.}\ \bibnamefont {Goerzen}},
  \bibinfo {author} {\bibfnamefont {S.}~\bibnamefont {Heinze}},\ and\ \bibinfo
  {author} {\bibfnamefont {D.}~\bibnamefont {Li}},\ }\bibfield  {title}
  {\bibinfo {title} {{Long lifetimes of nanoscale skyrmions in the
  lithium-decorated van der Waals ferromagnet
  ${\mathrm{Fe}}_{3}{\mathrm{GeTe}}_{2}$}},\ }\href
  {https://doi.org/10.1103/2926-8rf3} {\bibfield  {journal} {\bibinfo
  {journal} {Phys. Rev. B}\ }\textbf {\bibinfo {volume} {113}},\ \bibinfo
  {pages} {104445} (\bibinfo {year} {2026})}\BibitemShut {NoStop}%
\bibitem [{\citenamefont {Heil}\ \emph {et~al.}(2019)\citenamefont {Heil},
  \citenamefont {Rosch},\ and\ \citenamefont {Masell}}]{Masell2019}%
  \BibitemOpen
  \bibfield  {author} {\bibinfo {author} {\bibfnamefont {B.}~\bibnamefont
  {Heil}}, \bibinfo {author} {\bibfnamefont {A.}~\bibnamefont {Rosch}},\ and\
  \bibinfo {author} {\bibfnamefont {J.}~\bibnamefont {Masell}},\ }\bibfield
  {title} {\bibinfo {title} {Universality of annihilation barriers of large
  magnetic skyrmions in chiral and frustrated magnets},\ }\href
  {https://doi.org/10.1103/PhysRevB.100.134424} {\bibfield  {journal} {\bibinfo
   {journal} {Phys. Rev. B}\ }\textbf {\bibinfo {volume} {100}},\ \bibinfo
  {pages} {134424} (\bibinfo {year} {2019})}\BibitemShut {NoStop}%
\bibitem [{\citenamefont {Okubo}\ \emph {et~al.}(2012)\citenamefont {Okubo},
  \citenamefont {Chung},\ and\ \citenamefont {Kawamura}}]{Okubo2012}%
  \BibitemOpen
  \bibfield  {author} {\bibinfo {author} {\bibfnamefont {T.}~\bibnamefont
  {Okubo}}, \bibinfo {author} {\bibfnamefont {S.}~\bibnamefont {Chung}},\ and\
  \bibinfo {author} {\bibfnamefont {H.}~\bibnamefont {Kawamura}},\ }\bibfield
  {title} {\bibinfo {title} {Multiple-$q$ states and the skyrmion lattice of
  the triangular-lattice {Heisenberg} antiferromagnet under magnetic fields},\
  }\href {https://doi.org/10.1103/PhysRevLett.108.017206} {\bibfield  {journal}
  {\bibinfo  {journal} {Phys. Rev. Lett.}\ }\textbf {\bibinfo {volume} {108}},\
  \bibinfo {pages} {017206} (\bibinfo {year} {2012})}\BibitemShut {NoStop}%
\bibitem [{\citenamefont {Leonov}\ and\ \citenamefont
  {Mostovoy}(2015)}]{leonov2015multiply}%
  \BibitemOpen
  \bibfield  {author} {\bibinfo {author} {\bibfnamefont {A.}~\bibnamefont
  {Leonov}}\ and\ \bibinfo {author} {\bibfnamefont {M.}~\bibnamefont
  {Mostovoy}},\ }\bibfield  {title} {\bibinfo {title} {Multiply periodic states
  and isolated skyrmions in an anisotropic frustrated magnet},\ }\href
  {https://doi.org/10.1038/ncomms9275} {\bibfield  {journal} {\bibinfo
  {journal} {Nat. Commun.}\ }\textbf {\bibinfo {volume} {6}},\ \bibinfo {pages}
  {8275} (\bibinfo {year} {2015})}\BibitemShut {NoStop}%
\bibitem [{\citenamefont {Dup{\'e}}\ \emph {et~al.}(2016)\citenamefont
  {Dup{\'e}}, \citenamefont {Bihlmayer}, \citenamefont {B{\"o}ttcher},
  \citenamefont {Bl{\"u}gel},\ and\ \citenamefont
  {Heinze}}]{dupe2016engineering}%
  \BibitemOpen
  \bibfield  {author} {\bibinfo {author} {\bibfnamefont {B.}~\bibnamefont
  {Dup{\'e}}}, \bibinfo {author} {\bibfnamefont {G.}~\bibnamefont {Bihlmayer}},
  \bibinfo {author} {\bibfnamefont {M.}~\bibnamefont {B{\"o}ttcher}}, \bibinfo
  {author} {\bibfnamefont {S.}~\bibnamefont {Bl{\"u}gel}},\ and\ \bibinfo
  {author} {\bibfnamefont {S.}~\bibnamefont {Heinze}},\ }\bibfield  {title}
  {\bibinfo {title} {Engineering skyrmions in transition-metal multilayers for
  spintronics},\ }\href {https://doi.org/10.1038/ncomms11779} {\bibfield
  {journal} {\bibinfo  {journal} {Nat. Commun.}\ }\textbf {\bibinfo {volume}
  {7}},\ \bibinfo {pages} {11779} (\bibinfo {year} {2016})}\BibitemShut
  {NoStop}%
\bibitem [{\citenamefont {Rybakov}\ \emph {et~al.}(2022)\citenamefont
  {Rybakov}, \citenamefont {Kiselev}, \citenamefont {Borisov}, \citenamefont
  {D{\"o}ring}, \citenamefont {Melcher},\ and\ \citenamefont
  {Bl{\"u}gel}}]{rybakov2022magnetic}%
  \BibitemOpen
  \bibfield  {author} {\bibinfo {author} {\bibfnamefont {F.~N.}\ \bibnamefont
  {Rybakov}}, \bibinfo {author} {\bibfnamefont {N.~S.}\ \bibnamefont
  {Kiselev}}, \bibinfo {author} {\bibfnamefont {A.~B.}\ \bibnamefont
  {Borisov}}, \bibinfo {author} {\bibfnamefont {L.}~\bibnamefont {D{\"o}ring}},
  \bibinfo {author} {\bibfnamefont {C.}~\bibnamefont {Melcher}},\ and\ \bibinfo
  {author} {\bibfnamefont {S.}~\bibnamefont {Bl{\"u}gel}},\ }\bibfield  {title}
  {\bibinfo {title} {Magnetic hopfions in solids},\ }\href
  {https://doi.org/10.1063/5.0099942} {\bibfield  {journal} {\bibinfo
  {journal} {APL Mater.}\ }\textbf {\bibinfo {volume} {10}},\ \bibinfo {pages}
  {111113} (\bibinfo {year} {2022})}\BibitemShut {NoStop}%
\bibitem [{\citenamefont {Sallermann}\ \emph {et~al.}(2023)\citenamefont
  {Sallermann}, \citenamefont {J{\'o}nsson},\ and\ \citenamefont
  {Bl{\"u}gel}}]{sallermann2023stability}%
  \BibitemOpen
  \bibfield  {author} {\bibinfo {author} {\bibfnamefont {M.}~\bibnamefont
  {Sallermann}}, \bibinfo {author} {\bibfnamefont {H.}~\bibnamefont
  {J{\'o}nsson}},\ and\ \bibinfo {author} {\bibfnamefont {S.}~\bibnamefont
  {Bl{\"u}gel}},\ }\bibfield  {title} {\bibinfo {title} {Stability of hopfions
  in bulk magnets with competing exchange interactions},\ }\href
  {https://doi.org/10.1103/PhysRevB.107.104404} {\bibfield  {journal} {\bibinfo
   {journal} {Phys. Rev. B}\ }\textbf {\bibinfo {volume} {107}},\ \bibinfo
  {pages} {104404} (\bibinfo {year} {2023})}\BibitemShut {NoStop}%
\bibitem [{\citenamefont {Yosida}(1996)}]{Yosida1996}%
  \BibitemOpen
  \bibfield  {author} {\bibinfo {author} {\bibfnamefont {K.}~\bibnamefont
  {Yosida}},\ }\href {https://doi.org/10.1007/978-3-642-58900-3} {\emph
  {\bibinfo {title} {Theory of Magnetism}}},\ Springer Series in Solid-State
  Sciences\ (\bibinfo  {publisher} {Springer},\ \bibinfo {address} {Berlin,
  Heidelberg},\ \bibinfo {year} {1996})\BibitemShut {NoStop}%
\bibitem [{\citenamefont {Litvinov}\ and\ \citenamefont
  {Dugaev}(1998)}]{Lit1998}%
  \BibitemOpen
  \bibfield  {author} {\bibinfo {author} {\bibfnamefont {V.~I.}\ \bibnamefont
  {Litvinov}}\ and\ \bibinfo {author} {\bibfnamefont {V.~K.}\ \bibnamefont
  {Dugaev}},\ }\bibfield  {title} {\bibinfo {title} {{RKKY} interaction in one-
  and two-dimensional electron gases},\ }\href
  {https://doi.org/10.1103/PhysRevB.58.3584} {\bibfield  {journal} {\bibinfo
  {journal} {Phys. Rev. B}\ }\textbf {\bibinfo {volume} {58}},\ \bibinfo
  {pages} {3584} (\bibinfo {year} {1998})}\BibitemShut {NoStop}%
\bibitem [{\citenamefont {Bruno}(1995)}]{Bruno1995}%
  \BibitemOpen
  \bibfield  {author} {\bibinfo {author} {\bibfnamefont {P.}~\bibnamefont
  {Bruno}},\ }\bibfield  {title} {\bibinfo {title} {Theory of interlayer
  magnetic coupling},\ }\href {https://doi.org/10.1103/PhysRevB.52.411}
  {\bibfield  {journal} {\bibinfo  {journal} {Phys. Rev. B}\ }\textbf {\bibinfo
  {volume} {52}},\ \bibinfo {pages} {411} (\bibinfo {year} {1995})}\BibitemShut
  {NoStop}%
\bibitem [{\citenamefont {Hayami}\ \emph {et~al.}(2016)\citenamefont {Hayami},
  \citenamefont {Lin},\ and\ \citenamefont {Batista}}]{Hayami2016}%
  \BibitemOpen
  \bibfield  {author} {\bibinfo {author} {\bibfnamefont {S.}~\bibnamefont
  {Hayami}}, \bibinfo {author} {\bibfnamefont {S.-Z.}\ \bibnamefont {Lin}},\
  and\ \bibinfo {author} {\bibfnamefont {C.~D.}\ \bibnamefont {Batista}},\
  }\bibfield  {title} {\bibinfo {title} {Bubble and skyrmion crystals in
  frustrated magnets with easy-axis anisotropy},\ }\href
  {https://doi.org/10.1103/PhysRevB.93.184413} {\bibfield  {journal} {\bibinfo
  {journal} {Phys. Rev. B}\ }\textbf {\bibinfo {volume} {93}},\ \bibinfo
  {pages} {184413} (\bibinfo {year} {2016})}\BibitemShut {NoStop}%
\bibitem [{\citenamefont {Zhang}\ \emph {et~al.}(2017)\citenamefont {Zhang},
  \citenamefont {Xia}, \citenamefont {Zhou}, \citenamefont {Liu}, \citenamefont
  {Zhang},\ and\ \citenamefont {Ezawa}}]{zhang2017skyrmion}%
  \BibitemOpen
  \bibfield  {author} {\bibinfo {author} {\bibfnamefont {X.}~\bibnamefont
  {Zhang}}, \bibinfo {author} {\bibfnamefont {J.}~\bibnamefont {Xia}}, \bibinfo
  {author} {\bibfnamefont {Y.}~\bibnamefont {Zhou}}, \bibinfo {author}
  {\bibfnamefont {X.}~\bibnamefont {Liu}}, \bibinfo {author} {\bibfnamefont
  {H.}~\bibnamefont {Zhang}},\ and\ \bibinfo {author} {\bibfnamefont
  {M.}~\bibnamefont {Ezawa}},\ }\bibfield  {title} {\bibinfo {title} {Skyrmion
  dynamics in a frustrated ferromagnetic film and current-induced helicity
  locking-unlocking transition},\ }\href
  {https://doi.org/10.1038/s41467-017-01785-w} {\bibfield  {journal} {\bibinfo
  {journal} {Nat. Commun.}\ }\textbf {\bibinfo {volume} {8}},\ \bibinfo {pages}
  {1717} (\bibinfo {year} {2017})}\BibitemShut {NoStop}%
\bibitem [{\citenamefont {Desplat}\ \emph {et~al.}(2019)\citenamefont
  {Desplat}, \citenamefont {Kim},\ and\ \citenamefont
  {Stamps}}]{desplat2019paths}%
  \BibitemOpen
  \bibfield  {author} {\bibinfo {author} {\bibfnamefont {L.}~\bibnamefont
  {Desplat}}, \bibinfo {author} {\bibfnamefont {J.-V.}\ \bibnamefont {Kim}},\
  and\ \bibinfo {author} {\bibfnamefont {R.~L.}\ \bibnamefont {Stamps}},\
  }\bibfield  {title} {\bibinfo {title} {Paths to annihilation of first- and
  second-order (anti)skyrmions via (anti)meron nucleation on the frustrated
  square lattice},\ }\href {https://doi.org/10.1103/PhysRevB.99.174409}
  {\bibfield  {journal} {\bibinfo  {journal} {Phys. Rev. B}\ }\textbf {\bibinfo
  {volume} {99}},\ \bibinfo {pages} {174409} (\bibinfo {year}
  {2019})}\BibitemShut {NoStop}%
\bibitem [{\citenamefont {Zhang}\ \emph {et~al.}(2021)\citenamefont {Zhang},
  \citenamefont {Xia}, \citenamefont {Tretiakov}, \citenamefont {Diep},
  \citenamefont {Zhao}, \citenamefont {Yang}, \citenamefont {Zhou},
  \citenamefont {Ezawa},\ and\ \citenamefont {Liu}}]{Zhang2021}%
  \BibitemOpen
  \bibfield  {author} {\bibinfo {author} {\bibfnamefont {X.}~\bibnamefont
  {Zhang}}, \bibinfo {author} {\bibfnamefont {J.}~\bibnamefont {Xia}}, \bibinfo
  {author} {\bibfnamefont {O.~A.}\ \bibnamefont {Tretiakov}}, \bibinfo {author}
  {\bibfnamefont {H.~T.}\ \bibnamefont {Diep}}, \bibinfo {author}
  {\bibfnamefont {G.}~\bibnamefont {Zhao}}, \bibinfo {author} {\bibfnamefont
  {J.}~\bibnamefont {Yang}}, \bibinfo {author} {\bibfnamefont {Y.}~\bibnamefont
  {Zhou}}, \bibinfo {author} {\bibfnamefont {M.}~\bibnamefont {Ezawa}},\ and\
  \bibinfo {author} {\bibfnamefont {X.}~\bibnamefont {Liu}},\ }\bibfield
  {title} {\bibinfo {title} {Dynamic transformation between a skyrmion string
  and a bimeron string in a layered frustrated system},\ }\href
  {https://doi.org/10.1103/PhysRevB.104.L220406} {\bibfield  {journal}
  {\bibinfo  {journal} {Phys. Rev. B}\ }\textbf {\bibinfo {volume} {104}},\
  \bibinfo {pages} {L220406} (\bibinfo {year} {2021})}\BibitemShut {NoStop}%
\bibitem [{\citenamefont {Bessarab}\ \emph {et~al.}(2015)\citenamefont
  {Bessarab}, \citenamefont {Uzdin},\ and\ \citenamefont
  {J{\'o}nsson}}]{bessarab2015method}%
  \BibitemOpen
  \bibfield  {author} {\bibinfo {author} {\bibfnamefont {P.~F.}\ \bibnamefont
  {Bessarab}}, \bibinfo {author} {\bibfnamefont {V.~M.}\ \bibnamefont
  {Uzdin}},\ and\ \bibinfo {author} {\bibfnamefont {H.}~\bibnamefont
  {J{\'o}nsson}},\ }\bibfield  {title} {\bibinfo {title} {Method for finding
  mechanism and activation energy of magnetic transitions, applied to skyrmion
  and antivortex annihilation},\ }\href
  {https://doi.org/10.1016/j.cpc.2015.07.001} {\bibfield  {journal} {\bibinfo
  {journal} {Comput. Phys. Commun.}\ }\textbf {\bibinfo {volume} {196}},\
  \bibinfo {pages} {335} (\bibinfo {year} {2015})}\BibitemShut {NoStop}%
\bibitem [{sup()}]{supplmat}%
  \BibitemOpen
  \href@noop {} {}\bibinfo {note} {See Supplemental Material at
  http://link.aps.org/supplemental/ for micromagnetic model with fourth-order
  terms, effect of symmetry: square versus hexagonal lattices, representative
  parameter space of {RKKY}-type interactions, solving exchange coupling at
  fixed micromagnetic parameters, geodesic nudged elastic band ({GNEB}) method,
  minimal model for frustration range transition, exchange optimization under
  physical constraints, fitting and parameterization, contrast between skyrmion
  and saddle point projections, role of lattice symmetry for energy barrier
  optimization, and role of micromagnetic constraints in the optimization which
  includes
  Refs.~\cite{rybakov2022magnetic,sallermann2023stability,cranmer2023interpretable}.}\BibitemShut
  {Stop}%
\bibitem [{\citenamefont {Banik}\ \emph {et~al.}(2026)\citenamefont {Banik},
  \citenamefont {Kiselev},\ and\ \citenamefont {Nandy}}]{banik2026paradoxical}%
  \BibitemOpen
  \bibfield  {author} {\bibinfo {author} {\bibfnamefont {S.}~\bibnamefont
  {Banik}}, \bibinfo {author} {\bibfnamefont {N.~S.}\ \bibnamefont {Kiselev}},\
  and\ \bibinfo {author} {\bibfnamefont {A.~K.}\ \bibnamefont {Nandy}},\
  }\bibfield  {title} {\bibinfo {title} {Paradoxical topological soliton
  lattice in anisotropic frustrated chiral magnets},\ }\href
  {https://doi.org/10.1002/advs.202514568} {\bibfield  {journal} {\bibinfo
  {journal} {Adv. Sci.}\ }\textbf {\bibinfo {volume} {13}},\ \bibinfo {pages}
  {e14568} (\bibinfo {year} {2026})}\BibitemShut {NoStop}%
\bibitem [{\citenamefont {Polyakov}\ and\ \citenamefont
  {Belavin}(1975)}]{polyakov22metastable}%
  \BibitemOpen
  \bibfield  {author} {\bibinfo {author} {\bibfnamefont {A.}~\bibnamefont
  {Polyakov}}\ and\ \bibinfo {author} {\bibfnamefont {A.}~\bibnamefont
  {Belavin}},\ }\bibfield  {title} {\bibinfo {title} {Metastable states of
  two-dimensional isotropic ferromagnets},\ }\href
  {https://www.cpt.univ-mrs.fr/~verga/pdfs/Belavin-1975xw.pdf} {\bibfield
  {journal} {\bibinfo  {journal} {JETP Lett.}\ }\textbf {\bibinfo {volume}
  {22}},\ \bibinfo {pages} {245} (\bibinfo {year} {1975})}\BibitemShut
  {NoStop}%
\bibitem [{\citenamefont {B{\"u}ttner}\ \emph {et~al.}(2018)\citenamefont
  {B{\"u}ttner}, \citenamefont {Lemesh},\ and\ \citenamefont
  {Beach}}]{buttner2018theory}%
  \BibitemOpen
  \bibfield  {author} {\bibinfo {author} {\bibfnamefont {F.}~\bibnamefont
  {B{\"u}ttner}}, \bibinfo {author} {\bibfnamefont {I.}~\bibnamefont
  {Lemesh}},\ and\ \bibinfo {author} {\bibfnamefont {G.~S.}\ \bibnamefont
  {Beach}},\ }\bibfield  {title} {\bibinfo {title} {Theory of isolated magnetic
  skyrmions: From fundamentals to room temperature applications},\ }\href
  {https://doi.org/10.1038/s41598-018-22242-8} {\bibfield  {journal} {\bibinfo
  {journal} {Sci. Rep.}\ }\textbf {\bibinfo {volume} {8}},\ \bibinfo {pages}
  {4464} (\bibinfo {year} {2018})}\BibitemShut {NoStop}%
\bibitem [{\citenamefont {Bernand-Mantel}\ \emph {et~al.}(2022)\citenamefont
  {Bernand-Mantel}, \citenamefont {Muratov},\ and\ \citenamefont
  {Slastikov}}]{bernand2022micromagnetic}%
  \BibitemOpen
  \bibfield  {author} {\bibinfo {author} {\bibfnamefont {A.}~\bibnamefont
  {Bernand-Mantel}}, \bibinfo {author} {\bibfnamefont {C.~B.}\ \bibnamefont
  {Muratov}},\ and\ \bibinfo {author} {\bibfnamefont {V.~V.}\ \bibnamefont
  {Slastikov}},\ }\bibfield  {title} {\bibinfo {title} {A micromagnetic theory
  of skyrmion lifetime in ultrathin ferromagnetic films},\ }\href
  {https://doi.org/10.1073/pnas.2122237119} {\bibfield  {journal} {\bibinfo
  {journal} {Proc. Natl. Acad. Sci. U.S.A.}\ }\textbf {\bibinfo {volume}
  {119}},\ \bibinfo {pages} {e2122237119} (\bibinfo {year} {2022})}\BibitemShut
  {NoStop}%
\bibitem [{\citenamefont {Bernand-Mantel}\ \emph {et~al.}(2023)\citenamefont
  {Bernand-Mantel}, \citenamefont {Fondet}, \citenamefont {Barnova},
  \citenamefont {Simon},\ and\ \citenamefont {Muratov}}]{bernand2023theory}%
  \BibitemOpen
  \bibfield  {author} {\bibinfo {author} {\bibfnamefont {A.}~\bibnamefont
  {Bernand-Mantel}}, \bibinfo {author} {\bibfnamefont {A.}~\bibnamefont
  {Fondet}}, \bibinfo {author} {\bibfnamefont {S.}~\bibnamefont {Barnova}},
  \bibinfo {author} {\bibfnamefont {T.~M.}\ \bibnamefont {Simon}},\ and\
  \bibinfo {author} {\bibfnamefont {C.~B.}\ \bibnamefont {Muratov}},\
  }\bibfield  {title} {\bibinfo {title} {Theory of magnetic field stabilized
  compact skyrmions in thin-film ferromagnets},\ }\href
  {https://doi.org/10.1103/PhysRevB.108.L161405} {\bibfield  {journal}
  {\bibinfo  {journal} {Phys. Rev. B}\ }\textbf {\bibinfo {volume} {108}},\
  \bibinfo {pages} {L161405} (\bibinfo {year} {2023})}\BibitemShut {NoStop}%
\bibitem [{\citenamefont {Goerzen}\ \emph {et~al.}(2023)\citenamefont
  {Goerzen}, \citenamefont {von Malottki}, \citenamefont {Meyer}, \citenamefont
  {Bessarab},\ and\ \citenamefont {Heinze}}]{goerzen2023lifetime}%
  \BibitemOpen
  \bibfield  {author} {\bibinfo {author} {\bibfnamefont {M.~A.}\ \bibnamefont
  {Goerzen}}, \bibinfo {author} {\bibfnamefont {S.}~\bibnamefont {von
  Malottki}}, \bibinfo {author} {\bibfnamefont {S.}~\bibnamefont {Meyer}},
  \bibinfo {author} {\bibfnamefont {P.~F.}\ \bibnamefont {Bessarab}},\ and\
  \bibinfo {author} {\bibfnamefont {S.}~\bibnamefont {Heinze}},\ }\bibfield
  {title} {\bibinfo {title} {Lifetime of coexisting sub-10 nm zero-field
  skyrmions and antiskyrmions},\ }\href
  {https://doi.org/10.1038/s41535-023-00586-3} {\bibfield  {journal} {\bibinfo
  {journal} {npj Quantum Mater.}\ }\textbf {\bibinfo {volume} {8}},\ \bibinfo
  {pages} {54} (\bibinfo {year} {2023})}\BibitemShut {NoStop}%
\bibitem [{\citenamefont {Goerzen}\ \emph {et~al.}(2026)\citenamefont
  {Goerzen}, \citenamefont {Drevelow}, \citenamefont {Schrautzer},
  \citenamefont {Haldar}, \citenamefont {Heinze},\ and\ \citenamefont
  {Li}}]{Moritz_prb2026}%
  \BibitemOpen
  \bibfield  {author} {\bibinfo {author} {\bibfnamefont {M.~A.}\ \bibnamefont
  {Goerzen}}, \bibinfo {author} {\bibfnamefont {T.}~\bibnamefont {Drevelow}},
  \bibinfo {author} {\bibfnamefont {H.}~\bibnamefont {Schrautzer}}, \bibinfo
  {author} {\bibfnamefont {S.}~\bibnamefont {Haldar}}, \bibinfo {author}
  {\bibfnamefont {S.}~\bibnamefont {Heinze}},\ and\ \bibinfo {author}
  {\bibfnamefont {D.}~\bibnamefont {Li}},\ }\bibfield  {title} {\bibinfo
  {title} {Emergence of multiple topological spin textures in an all-magnetic
  van der waals heterostructure},\ }\href {https://doi.org/10.1103/g149-5gyd}
  {\bibfield  {journal} {\bibinfo  {journal} {Phys. Rev. B}\ }\textbf {\bibinfo
  {volume} {113}},\ \bibinfo {pages} {094420} (\bibinfo {year}
  {2026})}\BibitemShut {NoStop}%
\bibitem [{\citenamefont {Zhu}\ \emph {et~al.}(2026)\citenamefont {Zhu},
  \citenamefont {Goerzen}, \citenamefont {Song}, \citenamefont {Heinze},\ and\
  \citenamefont {Li}}]{shiwei2026}%
  \BibitemOpen
  \bibfield  {author} {\bibinfo {author} {\bibfnamefont {S.}~\bibnamefont
  {Zhu}}, \bibinfo {author} {\bibfnamefont {M.~A.}\ \bibnamefont {Goerzen}},
  \bibinfo {author} {\bibfnamefont {C.}~\bibnamefont {Song}}, \bibinfo {author}
  {\bibfnamefont {S.}~\bibnamefont {Heinze}},\ and\ \bibinfo {author}
  {\bibfnamefont {D.}~\bibnamefont {Li}},\ }\bibfield  {title} {\bibinfo
  {title} {Strongly enhanced lifetime of higher-order bimerons and
  antibimerons},\ }\href {https://doi.org/10.1021/acs.nanolett.6c00092}
  {\bibfield  {journal} {\bibinfo  {journal} {Nano Lett.}\ }\textbf {\bibinfo
  {volume} {26}},\ \bibinfo {pages} {6307} (\bibinfo {year}
  {2026})}\BibitemShut {NoStop}%
\bibitem [{\citenamefont {Zhang}\ \emph {et~al.}(2024)\citenamefont {Zhang},
  \citenamefont {Lu}, \citenamefont {Tabrizian}, \citenamefont {Feng},\ and\
  \citenamefont {Wu}}]{zhang20242d}%
  \BibitemOpen
  \bibfield  {author} {\bibinfo {author} {\bibfnamefont {B.}~\bibnamefont
  {Zhang}}, \bibinfo {author} {\bibfnamefont {P.}~\bibnamefont {Lu}}, \bibinfo
  {author} {\bibfnamefont {R.}~\bibnamefont {Tabrizian}}, \bibinfo {author}
  {\bibfnamefont {P.~X.-L.}\ \bibnamefont {Feng}},\ and\ \bibinfo {author}
  {\bibfnamefont {Y.}~\bibnamefont {Wu}},\ }\bibfield  {title} {\bibinfo
  {title} {{2D magnetic heterostructures: Spintronics and quantum future}},\
  }\href {https://doi.org/10.1038/s44306-024-00011-w} {\bibfield  {journal}
  {\bibinfo  {journal} {npj Spintronics}\ }\textbf {\bibinfo {volume} {2}},\
  \bibinfo {pages} {6} (\bibinfo {year} {2024})}\BibitemShut {NoStop}%
\bibitem [{\citenamefont {Cranmer}(2023)}]{cranmer2023interpretable}%
  \BibitemOpen
  \bibfield  {author} {\bibinfo {author} {\bibfnamefont {M.}~\bibnamefont
  {Cranmer}},\ }\bibfield  {title} {\bibinfo {title} {Interpretable machine
  learning for science with {PySR} and {S}ymbolic{R}egression. jl},\ }\href
  {https://arxiv.org/abs/2305.01582} {\bibfield  {journal} {\bibinfo  {journal}
  {arXiv preprint arXiv:2305.01582}\ } (\bibinfo {year} {2023})}\BibitemShut
  {NoStop}%
\end{thebibliography}%
    
\end{document}